\documentclass[11.5pt]{article}

\usepackage{amsmath}
\usepackage{natbib}
\usepackage[pdftex]{hyperref}
\usepackage{hypernat}
\usepackage{mathtools}
\usepackage{amsfonts}
\usepackage{dsfont}
\usepackage[T1]{fontenc}
\usepackage{graphicx,psfrag,epsf}
\usepackage{enumerate}
\usepackage{caption}
\usepackage[skip=0.5ex]{subcaption}
\usepackage{lscape}

\usepackage{fancyhdr}
\title{Estimating COVID-19 cases and reproduction number in Mexico}
\author{Michelle Anzarut\footnote{Correspondence: mich.anzarut@gmail.com} , Luis Felipe Gonz\'alez, Sonia Mendiz\'abal \\ and Mar\'ia Teresa Ortiz}

\date{July 16, 2020}

\begin{document}

\maketitle

\begin{abstract}
In this report we fit a semi-mechanistic Bayesian hierarchical model to describe the Mexican COVID-19 epidemic. We obtain two epidemiological measures: the number of infections and the reproduction number. Estimations are based on death data. Hence, we expect our estimates to be more accurate than the attack rates estimated from the reported number of cases.

\vspace*{5mm}


\end{abstract}

\section{Introduction}\label{section:Intro}

As of today, there are almost 14 million of confirmed cases of coronavirus disease 2019 (COVID-19) in the world, and a total of 591,228 reported deaths (\citeauthor{worldometers}). Mexico has 317,635 COVID-19 confirmed cases with 36,906 deaths. The numbers are growing rapidly in the country, along with plans for a gradual reopening which started in June. In addition to this, the testing rate in Mexico is the lowest among the OECD countries, implying a likely large sub-reporting of cases. The Ministry of Health of Mexico has declared that the data available are inconsistent and, therefore, difficult to use to define the reopening of activities. Given this, a better understanding of the current epidemiological situation at the state level, and the impact of government-imposed mitigation measures is necessary. 

\begin{figure}[p]
    \centering
    \includegraphics[scale = .17]{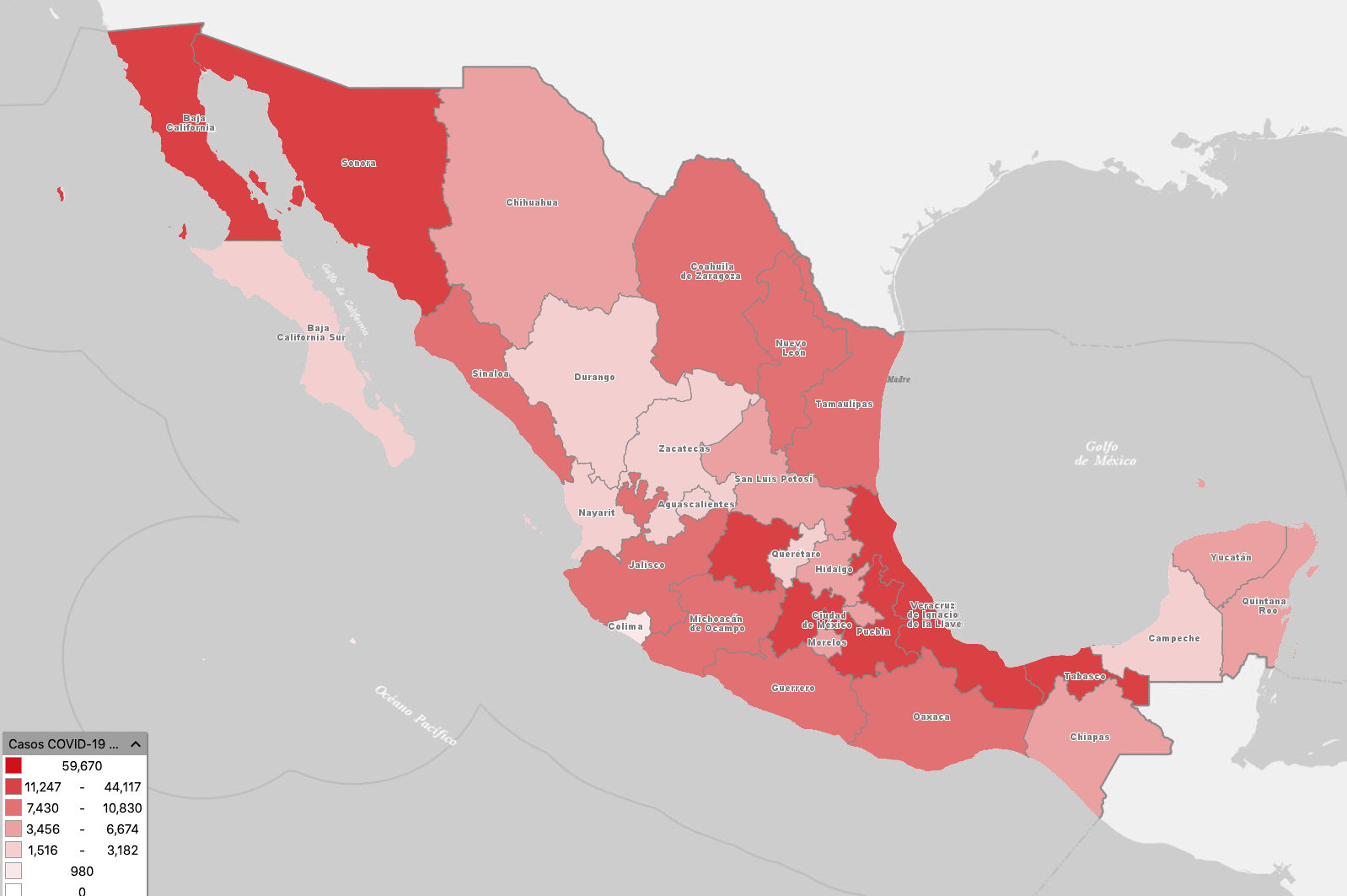}
     \includegraphics[scale = .17]{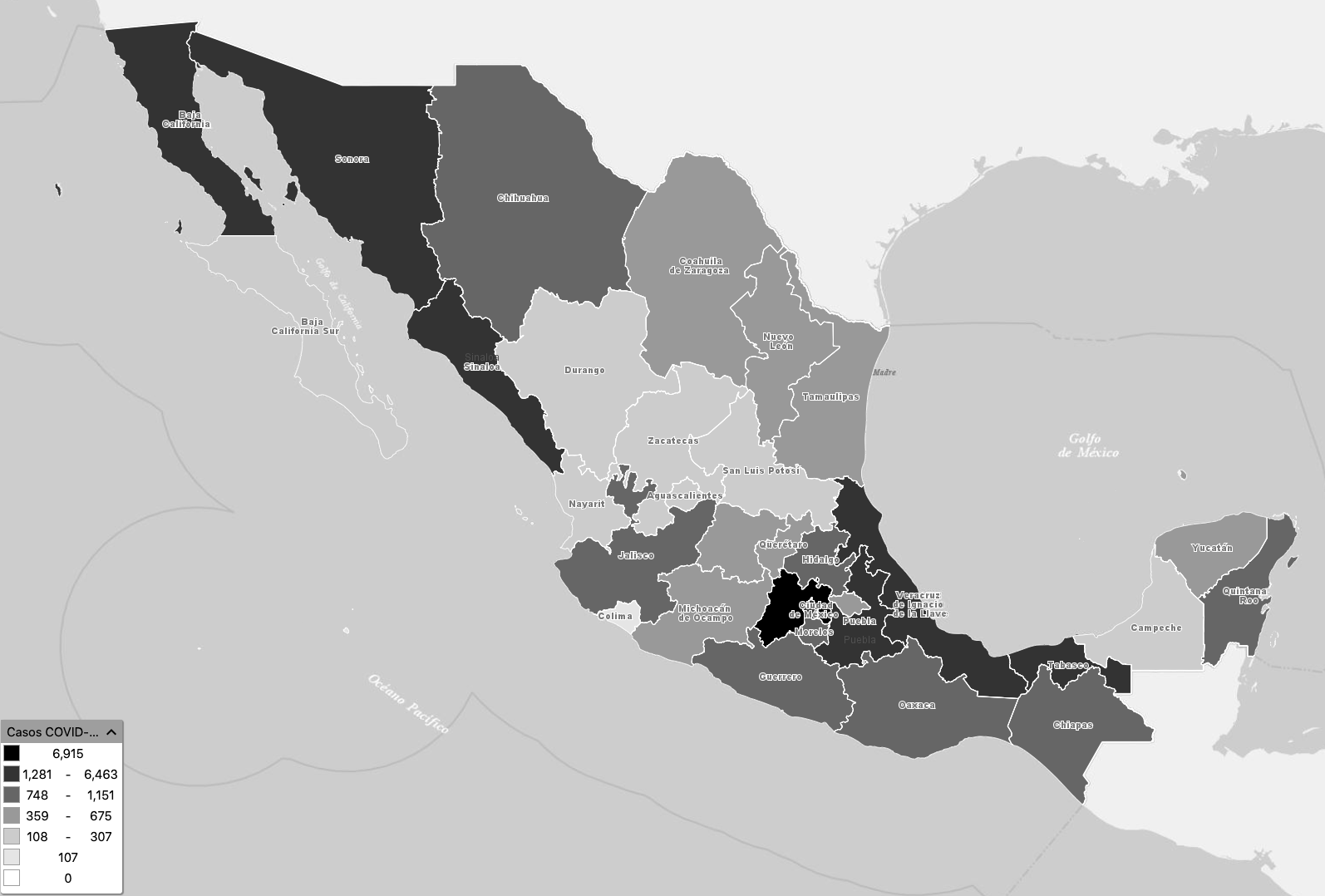}
    \caption{Map of COVID-19 cases (a) and deaths (b) in Mexico by state level. Data source: \citeauthor{mapas}}
    \label{fig:map}
\end{figure}

Here, we use the semi-mechanistic Bayesian hierarchical model of COVID-19 epidemiological dynamics found in \cite{flaxman2020report} to assess the effect of different control interventions in Mexico. We estimate the number of infections and the time-varying reproduction number (the expected number of secondary cases caused by each infected individual) as a function of human mobility. Figure \ref{fig:map} shows the cases and deaths reported in Mexico. The distribution of deaths among states is highly heterogeneous, with Mexico City and State of Mexico accounting for almost 40\% of the deaths reported to date. 

Mexico's government determined three phases of contingency for the COVID-19 pandemic. The first two cases were confirmed on February 27, from travellers returning from Italy. Later, on March 17, the suspension of school activities began nationally. A week later, on March 23, The Comit\'e Nacional para la Seguridad en Salud announced the start of the second phase of the epidemic. Almost a month later, on April 21, phase three was declared. Distancing measures included suspension of all non essential activities of public, private and social sectors. The measures were initially planned to last until April 30, but were later extended depending on the local situation of every municipality of the country. 

In June, the government announced a monitoring system to regulate the use of public space in accordance with the risk of COVID-19 contagion, with the aim of moving towards a new normality. They defined the traffic light of state epidemiological risk. This traffic light is made up of four colors: red, orange, yellow and green. Red is the color with the highest epidemiological risk, where only essential economic activities are allowed. The traffic light moves from color to color until reaching green, when all activities are allowed. In the last update, 17 entities were in orange, with high risk, and 15 in red, with maximum risk. It remains unclear the extent to which this traffic light has been effective. This week, the government did not present an epidemiological traffic light because state data is not consistent and the reopening has caused outbreaks. Given this, a better understanding of the current situation would be highly beneficial to guide policy decisions.

The outline of the paper is as follows. In Section \ref{section:Model} we explain briefly the model and some important modelling assumptions. In Section \ref{section:Data} we describe our data sources. In Section \ref{section:Results} we show the application of the model results. Finally, in Section \ref{section:Discussion} we give some concluding remarks and future research directions. 

\section{Model and Assumptions}\label{section:Model}

The main idea of  \cite{flaxman2020report} is to model the expected number of deaths in a given country on a given day as a function of the number of infections occurring in previous days, linking the deaths to the infected cases by using two quantities. The first is the infection fatality ratio (IFR), which is the probability of death given infection. The second is the distribution of times from infection to death. The number of deaths every day is the sum of the past infections weighted by their probability of death, where the probability of death depends on the number of days since infection and the state IFR.

For the state-level IFR, we made adjustments accounting for the substantial heterogeneity we expect to observe in health outcomes across states, due to variation in healthcare quality and capacity \citep[as in][]{mellan2020report}. We consider the IFR of Mexico City as the IFR estimated from the Chinese epidemic \citep[see][]{verity2020estimates}, and the IFR of Chiapas and Oaxaca, which are the most marginalized states in Mexico, more similar to those that would be expected in a Lower Middle Income Country \citep[see][]{walker2020report}. Then, we weight according to age groups. Finally, we interpolated the rest of the states using five levels of the \citeauthor{IM}. The marginalization index is an official standard indicator that allows states to be differentiated according to the deprivations suffered by the population, such as education, housing, monetary income, or access to health services. We also allow the IFR for every state to have additional noise, multiplying it by a $\mathit{N}(1, 0.1)$ distribution. 

The distribution of times from infection to death is the sum of two independent random times: the incubation period and the time between onset of symptoms and death. As in \cite{flaxman2020report}, we assume the infection-to-onset distribution is a $\mathit{Gamma}(5.1, 0.86)$, and the onset-to-death distribution is a $\mathit{Gamma}(18.8, 0.45)$. 

Now, in Mexico, there are two evident and significant problems that must be considered with death data. One of them is the death-delayed reporting and the other is the under-reporting. The death-delayed reporting refers to COVID-19 deaths that will take time to report (see \citeauthor{Garrido}). Death records must be validated by two information systems, which obtain information from medical units and health jurisdictions. Many of the COVID-19 deaths that were tested are awaiting the proper certification and classification. The assessment and confirmation processes take days. That is why the official figures of deaths from COVID-19, published by the Ministry of Health, have an important lag. 

\begin{figure}[p]
    \centering
    \includegraphics[scale = .39]{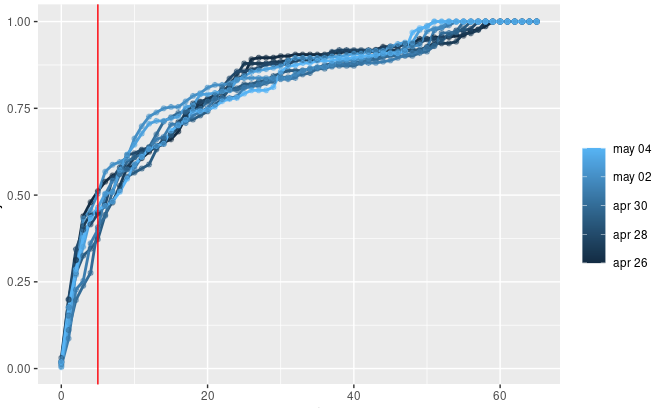}
    \caption{Cumulative proportion of deaths reported. The x-axis is the days that have passed since the deaths with different colours corresponding to different dates of death. }
    \label{fig:reporting_lag}
\end{figure}

Figure  \ref{fig:reporting_lag}  illustrates the cumulative proportion of deaths reported as the delay increases, with different colours corresponding to different dates of death. Very few deaths (around 10\%) are reported within one day, and the vast majority of deaths (around 90\%) are reported after a month. To account for this, we are adjusting the deaths reported in the following way. We assume the delay has the same distribution in all the country, but this can be corrected when more data becomes available. Let $p_t$ be the proportion of cases reported on day $t$, then
\begin{align*}
	(p_1,...,p_k) \sim Dir(\alpha\eta_1,...,\alpha\eta_{k}), 
\end{align*}
where the parameters $\eta_1,...,\eta_{k}$ are the empirical proportion of cases reported, and $\alpha \sim Gamma(100, 1)$. Notice that, since $\eta_1+...+\eta_{k}=1$, then $\mathbb{E}[p_t] = \eta_t$ and $Var(p_t) = \eta_t(1-\eta_t)(\alpha+1)^{-1}$. Afterwards, we are multiplying the case numbers of each day by the estimated proportion of cases reported. We do not report the last two days, since underreporting in those two days is very high. Also, to avoid biasing the model, we only include observed deaths from the day after a state has accumulated 10 deaths.  

The under-reporting refers to COVID-19 deaths that will not be reported (or it will take more than a year). Estimating the extent of the under-reporting remains very difficult, especially in areas outside of Mexico City where less studies have been conducted. Estimates of excess deaths can provide information about the burden of mortality potentially related to the COVID-19 pandemic, including deaths that are directly or indirectly attributed to COVID-19. However, in Mexico, there are no official published data on deaths during 2020. The publication of death statistics is annual and the last record is from 2018. 

In Mexico City, there are several death certificates stating that the confirmed or probable cause of death was COVID-19, three times higher than the COVID-19 deaths reported by the government (see \citeauthor{mexicanos_corrupc}). Another study of the death certificates shows that the excess mortality derived from the COVID-19 health crisis in Mexico City is four times greater than that reported (\citeauthor{actas_def}). This excess includes both people who died from COVID-19 and also those who died from other causes derived from the health crisis. According to the Registro Nacional de Poblaci\'on (Renapo), the number of deaths from COVID-19 in the country doubles what is reported every day (\citeauthor{RENAPO}). However, the government rejects that these death figures are official.

\begin{figure}[p]
    \centering
    \includegraphics[scale = .3]{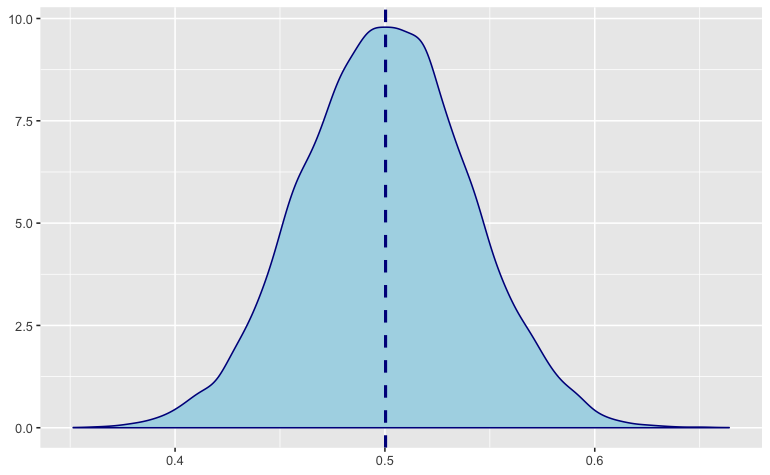}
    \caption{Prior distribution to death under-reporting, $\mathit{Beta}(80, 80)$.}
    \label{fig:beta}
\end{figure}

We adopted the extension of  \cite{mellan2020report} to reflect the uncertainty about under-reported deaths. Meaning that we address the effect of under-reporting in the data by setting a $\psi_m \sim \mathit{Beta}(\theta, \rho)$ prior distribution to death under-reporting in each state $m$, where both of the hyperparameters of the beta density are fixed in 80, in order to get a mode of 50\% (meaning that, on average, half of the COVID-19 deaths are being reported). Figure \ref{fig:beta} shows the plot of this prior distribution.

With all of this considerations, if we let $D_{t,m}$ be the daily deaths attributed to COVID-19 at day $t$ and state $m$, we model $D_{t,m}$ using a positive real-valued function $d_{t,m} = \mathbb{E}[\psi_m p_{t} D_{t,m}]$ that represents the expected number of deaths, taking into account the designated state-under-reported rate $\psi_m$ and the death-delayed reporting adjustment $p_t$. 

The rest of the mathematical model follows the original manuscript of \cite{flaxman2020report}. Daily deaths are assumed to follow a negative binomial distribution with mean $d_{t,m}$ and variance $d_{t,m} + \frac{d_{t,m}}{\phi}$ , where $\phi$ follows a positive half normal distribution, $\phi \sim \mathit{N}^{+}(0,5)$, and $d_{t,m}$ is given by the following discrete sum:
\begin{align*}
	d_{t,m} = \text{IFR}_m \sum_{\tau = 0}^{t-1} c_{\tau,m}\pi_{t-\tau}.
\end{align*}
$\text{IFR}_m$ is the IFR of the state $m$, $c_{\tau,m}$ is the number of new infections on day $\tau$ in state $m$, and $\pi$ is the discretized distribution of times from infection to death. 

To model the number of infections over time, we need to specify a serial interval distribution $g$ with density $g(\tau)$ (the time between when a person gets infected and when they subsequently infect another other people), which we set as in \cite{flaxman2020report}, $g \sim \mathit{Gamma}(6.5, 0.62)$. With this, the number of infections $c_{t,m}$ on a given day $t$, and a state $m$, is then given by:
\begin{align*}
	c_{t,m} = S_{t,m} R_{t,m} \sum_{\tau = 0}^{t-1} c_{\tau,m}g_{t-\tau},
\end{align*}
where $S_{t,m} = 1-N_m^{-1}\sum_{i=1}^{t-1} c_{i,m}$ is a population adjustment to account for susceptible individuals, $g_m$ is the discretized serial interval distribution, and $R_{t,m}$ is the state-specific time-varying reproduction number which was parametrized as a function of \citeauthor{movility_data} as explained below.

Denote by $X_{k,t,m}$ to the k-th Google mobility indicator at time $t$ for state $m$, smoothed by a seven day moving average (meaning that each point is the mean of the day and the previous 6 days). To allow flexibility, we define dummy variables after June, were relaxations started. Denote this variables by $Z_{k,t,m}$, then $Z_{k,t,m} = 1$ if the day $t$ is after June 1 or $Z_{k,t,m} = 0$ otherwise, for $k=1,...4$ and for any state $m$.  The state-specific time-varying reproduction number $R_{t,m}$ is modelled by:
\begin{align*}
	R_{t,m} = R_{0,m} \ 2\lambda^{-1}\left\lbrace - \sum_{k=1}^4 \left[ (\alpha_k + \beta_{k,m} ) X_{k,t,m} + \gamma_{k,m}Z_{k,t,m} \right] \right\rbrace,
\end{align*}

The basic reproduction number, $R_{0,m}$, has a normal prior for every state $m$, $R_{0,m} \sim \mathit{N}(3.28, k)$, where $k \sim \mathit{N}^{+}(0,0.5)$. This premise is tested in \cite{mellan2020report} through a sensitivity analysis. The conclusion is that, for this type of model, the results vary slightly depending on it. However, our major concern must be two highly influential assumptions, each state IFR and the under-reporting distribution.

\section{Data}\label{section:Data}

As input of deaths and reported cases, our model uses \citeauthor{daily_updates} from the Direcci\'on General de Epidemiolog\'ia, and the  \citeauthor{historical_bases}. For population counts we are using data provided by V\'ictor M. Garc\'ia as the Consejo Nacional de Poblaci\'on population projections for 2020 are not disaggregated for the age groups over 65 years old. Regarding intervention data, the values taken into account are the dates in which interventions were effectively applied, even though they were encouraged at earlier dates.

\section{Results}\label{section:Results}

The attack rate (AR) is the proportion of individuals that has been infected to date. The country's AR is estimated in 18\% (95\% Confidence Interval (CI): 16\%-19\%) on July 7. Of this infections, 7,235,600 (95\% CI: 6,281,900-8,062,000) have occurred in the previous 14 days (which is the official definition of active cases in Mexico).	

Table \ref{results} shows the estimated IFR, state population, reported deaths and deaths per million population, estimated number of infections in thousands, estimated number of infections in the previous 14 days in thousands, and estimated AR for all the states.

\begin{landscape}
\begin{table}[ht]
\caption{\label{results} Estimated results for all the states of Mexico as of July 7, 2020.}
\begin{small}
\begin{tabular}{| l | c c c c lll |}
  \hline
		State &   IFR \%  & Population & Deaths & \begin{tabular}{@{}c@{}} Deaths \\ per \\ million\end{tabular} & 
		\begin{tabular}{@{}c@{}} Infections \\ (thousands) \end{tabular} &
		\begin{tabular}{@{}c@{}} Infections \\ previous \\ 14 days \\ (thousands) \end{tabular} &
		Attack rate \% \\
		\hline
	State of Mexico & 0.59 & 17338220 & 6392 & 369 & 4460 [4070,4800] & 1130 [1020,1230] & 25.7 [23.5,27.7] \\ 
  Mexico City & 0.65 & 9025363 & 6119 & 678 & 3060 [2840,3270] & 473 [440,504] & 33.9 [31.5,36.2] \\ 
  Baja California & 0.40 & 3606940 & 2139 & 593 & 1710 [1590,1820] & 346 [322,367] & 47.4 [44.1,50.4] \\ 
  Veracruz & 1.00 & 8514724 & 1921 & 226 & 620 [561,672] & 90 [76,103] & 7.3 [6.6,7.9] \\ 
  Puebla & 0.87 & 6573843 & 1627 & 247 & 844 [750,925] & 254 [212,290] & 12.8 [11.4,14.1] \\ 
  Sinaloa & 0.80 & 3143980 & 1533 & 488 & 677 [616,731] & 142 [124,158] & 21.5 [19.6,23.3] \\ 
  Tabasco & 0.69 & 2558349 & 1276 & 499 & 857 [778,927] & 329 [292,362] & 33.5 [30.4,36.2] \\ 
  Sonora & 0.61 & 3056397 & 1069 & 350 & 1030 [925,1120] & 301 [273,327] & 33.6 [30.3,36.5] \\ 
  Guerrero & 1.00 & 3650850 & 1054 & 289 & 355 [321,385] & 67 [58,75] & 9.7 [8.8,10.5] \\ 
  Jalisco & 0.60 & 8368602 & 931 & 111 & 1190 [1070,1290] & 572 [507,630] & 14.2 [12.8,15.5] \\ 
  Chiapas & 0.84 & 5688998 & 728 & 128 & 362 [318,399] & 90 [74,104] & 6.4 [5.6,7] \\ 
  Morelos & 0.84 & 2033373 & 724 & 356 & 244 [217,268] & 28 [21,33] & 12 [10.7,13.2] \\ 
  Hidalgo & 0.93 & 3068696 & 723 & 236 & 362 [311,404] & 128 [101,149] & 11.8 [10.1,13.2] \\ 
  Chihuahua & 0.59 & 3783680 & 707 & 187 & 356 [314,391] & 50 [37,60] & 9.4 [8.3,10.3] \\ 
  Oaxaca & 1.10 & 4132318 & 695 & 168 & 288 [257,316] & 90 [78,101] & 7 [6.2,7.6] \\ 
  Quintana Roo & 0.58 & 1704010 & 666 & 391 & 663 [580,737] & 325 [275,370] & 38.9 [34,43.2] \\ 
  Michoac\'an & 0.91 & 4808791 & 551 & 115 & 243 [213,269] & 64 [51,73] & 5.1 [4.4,5.6] \\ 
  Yucat\'an & 0.94 & 2246505 & 545 & 243 & 463 [392,522] & 249 [201,290] & 20.6 [17.5,23.2] \\ 
  Tamaulipas & 0.63 & 3635833 & 528 & 145 & 574 [505,632] & 253 [217,284] & 15.8 [13.9,17.4] \\ 
  Guanajuato & 0.70 & 6201449 & 508 & 82 & 518 [450,577] & 244 [202,278] & 8.4 [7.3,9.3] \\ 
  Nuevo Le\'on & 0.45 & 5571904 & 487 & 87 & 933 [824,1030] & 499 [433,558] & 16.7 [14.8,18.5] \\ 
  Tlaxcala & 0.72 & 1372108 & 433 & 316 & 223 [195,247] & 47 [38,55] & 16.2 [14.2,18] \\ 
  Coahuila & 0.44 & 3197188 & 328 & 103 & 829 [733,912] & 509 [448,562] & 25.9 [22.9,28.5] \\ 
  Quer\'etaro & 0.53 & 2259471 & 325 & 144 & 485 [428,535] & 270 [236,300] & 21.4 [19,23.7] \\ 
  Campeche & 0.84 & 992306 & 258 & 260 & 176 [151,196] & 70 [58,80] & 17.7 [15.2,19.7] \\ 
  Nayarit & 0.76 & 1279671 & 233 & 182 & 223 [196,246] & 111 [97,123] & 17.4 [15.3,19.2] \\ 
  San Luis Potos\'i & 0.95 & 2856171 & 232 & 81 & 174 [144,198] & 84 [66,99] & 6.1 [5,6.9] \\ 
  Durango & 0.74 & 1861051 & 172 & 92 & 221 [187,249] & 131 [108,150] & 11.9 [10.1,13.4] \\ 
  Aguascalientes & 0.54 & 1425105 & 171 & 120 & 171 [139,197] & 63 [46,76] & 12 [9.7,13.8] \\ 
  Zacatecas & 0.77 & 1660543 & 129 & 78 & 159 [129,183] & 94 [73,111] & 9.6 [7.8,11] \\ 
  Baja California S & 0.53 & 796398 & 95 & 119 & 129 [97.6,154] & 70 [49,87] & 16.2 [12.3,19.3] \\ 
  Colima & 0.60 & 778989 & 82 & 105 & 113 [92.5,131] & 62 [50,72] & 14.5 [11.9,16.8] \\ 
   \hline
\end{tabular}
\end{small}
\end{table}
\end{landscape}

Our computed IFRs for each of the states range from 0.4\% to 1.1\%, reflecting substantial differences between states in their demographic structure and healthcare provision.  The percentage of people that have been infected with SARS-CoV-2 ranges from 47\% (95\% Cl: 44\%-50\%) in Baja California to 5\% (95\% Cl: 4\%-6\%) in Michoac\'an. 

Figure \ref{fig:infections} shows the estimates of infections, deaths and reproduction number for each state. The reproduction number is an important parameter for assessing whether additional interventions are required. At the start of the epidemic, the basic reproduction number meant that an infected individual would infect three others on average. If the reproduction number is above one, a sustained outbreak is likely. The aim of control interventions is typically to reduce the reproduction number below one. Following non-pharmaceutical interventions such as school closures and decreases in population mobility, our results show substantial reductions in the estimated value of $R_{t, m}$ in each state $m$. However, we can notice growth from June to this date, which has translated to  a linear growth of the infections. In some states, the number of infections goes down and then comes back up again, making what many call a second wave.

\begin{figure}[!b]
    \centering
    \begin{subfigure}{1.2\textwidth}
        \includegraphics[scale = .3]{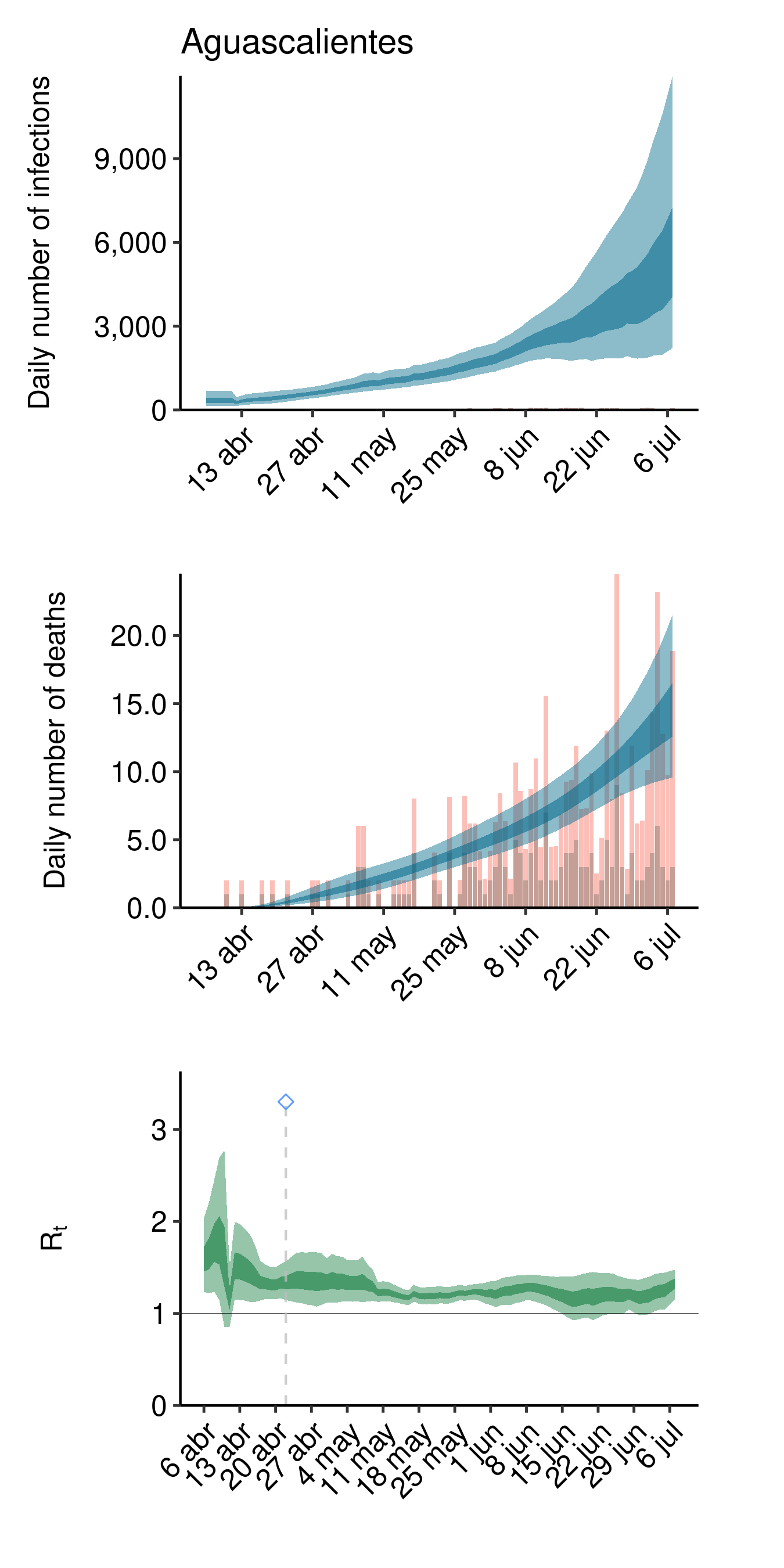}
        \includegraphics[scale = .3]{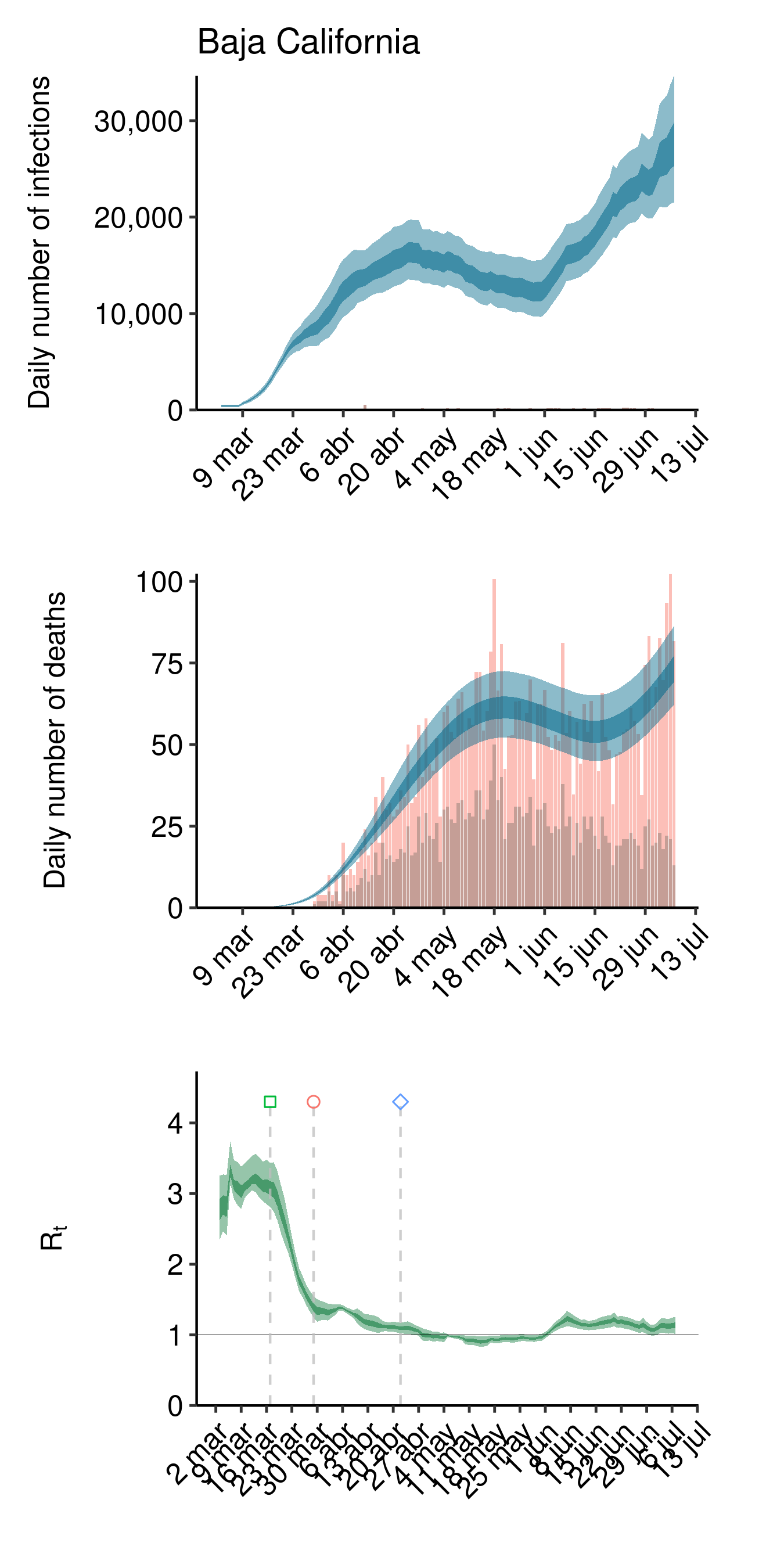}
         \includegraphics[scale = .3]{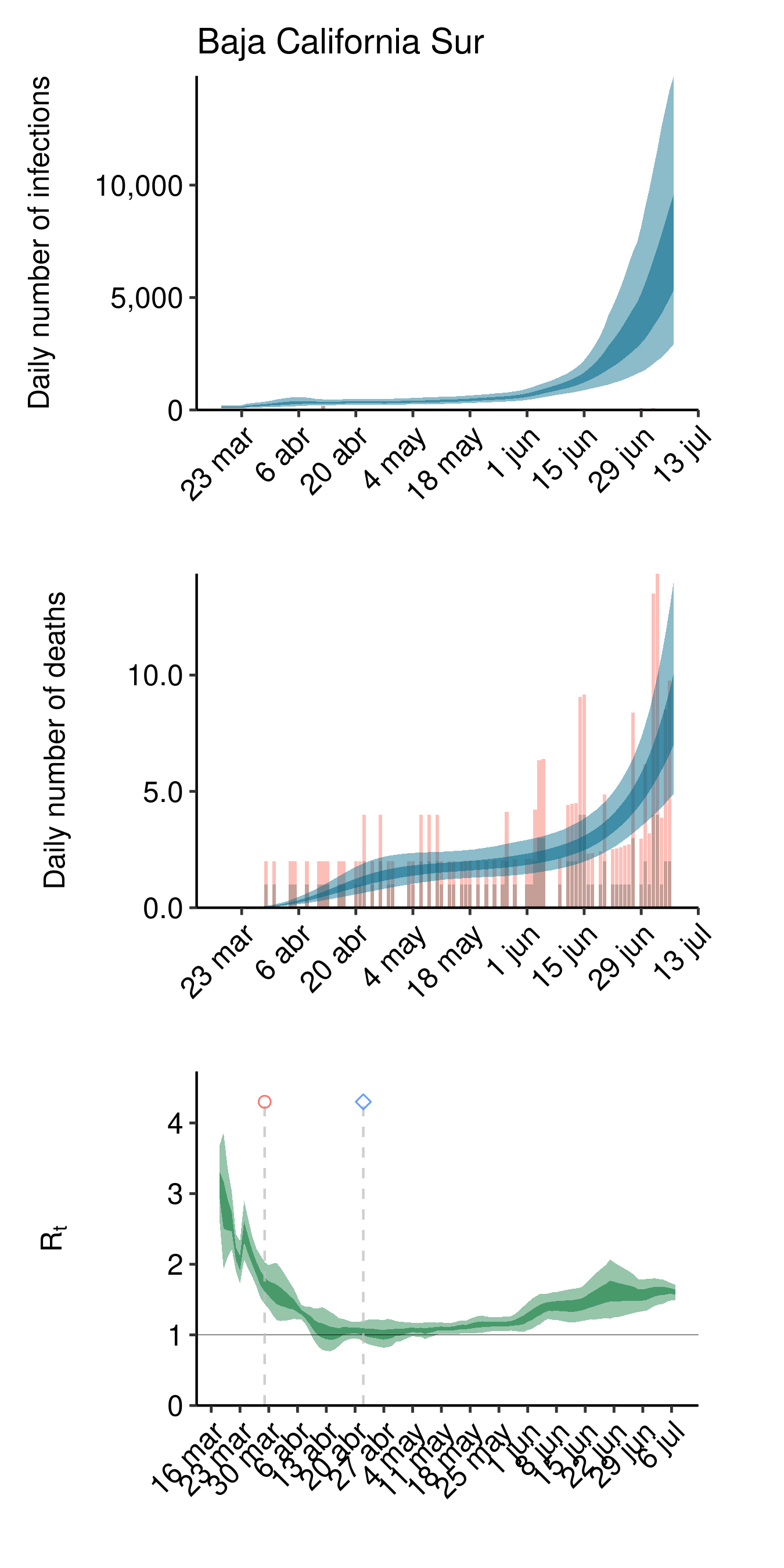}
          \includegraphics[scale = .3]{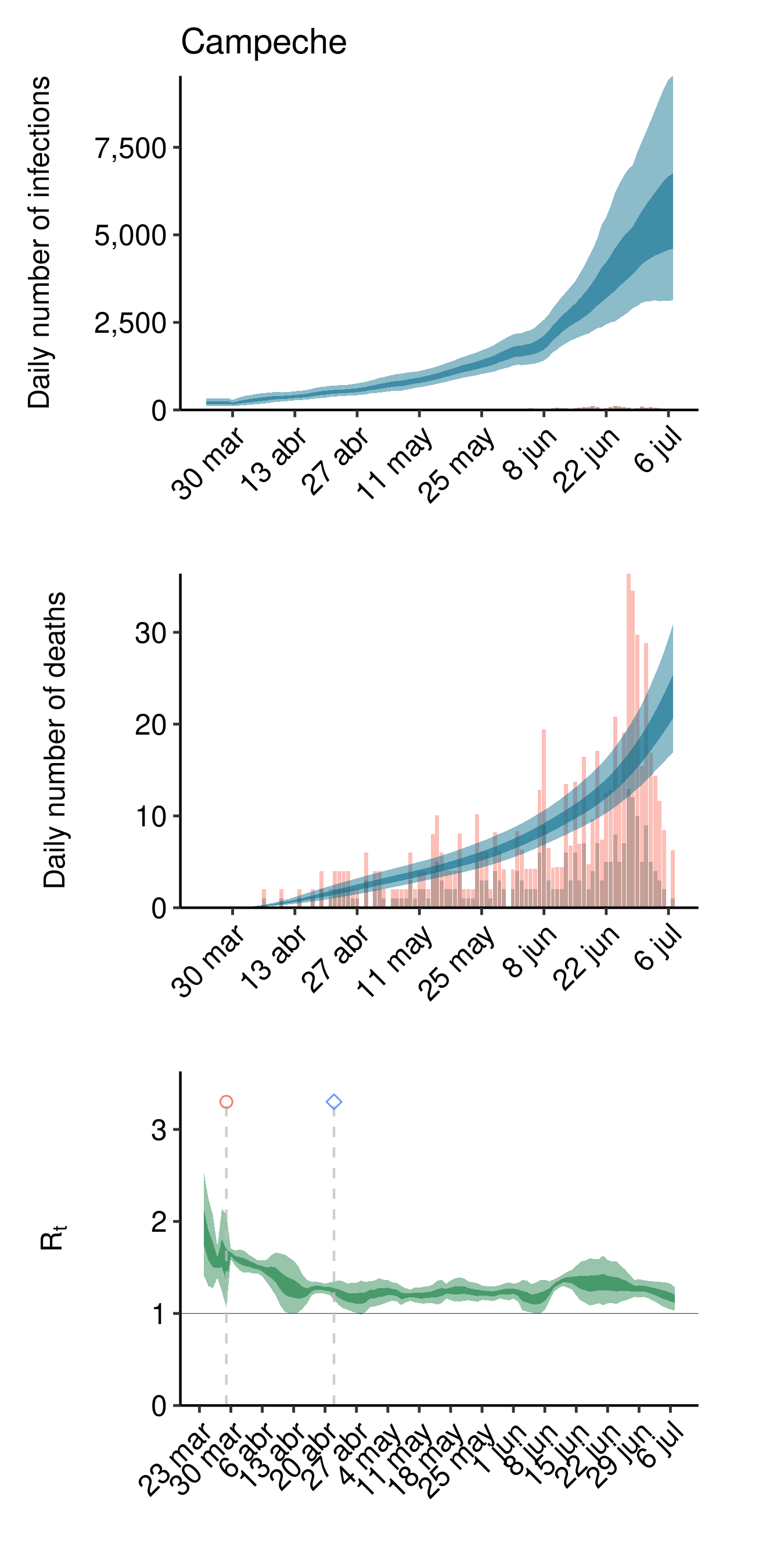}
    \end{subfigure}
	\medskip
    \begin{subfigure}{1.2\textwidth}
        \includegraphics[scale = .3]{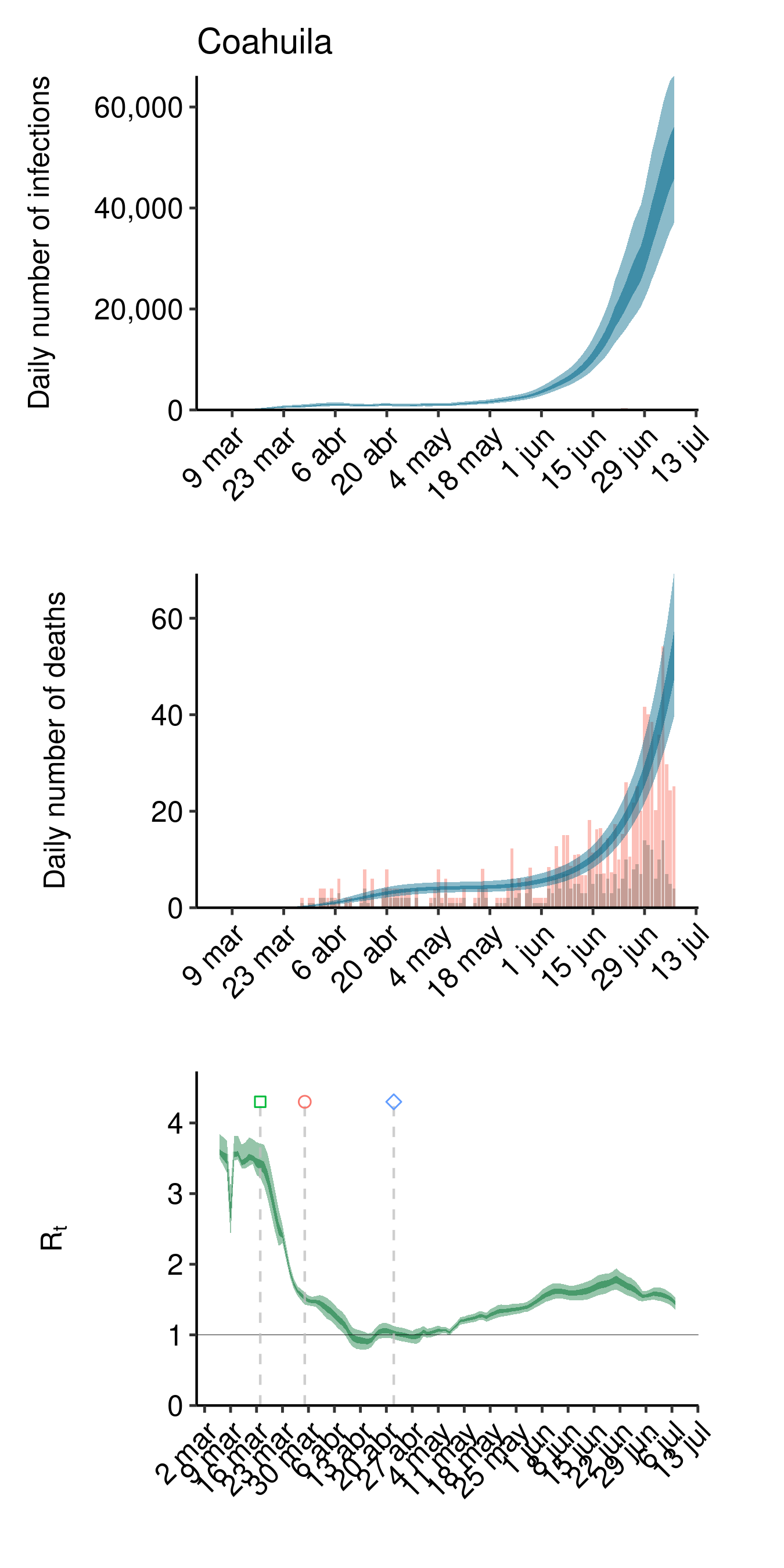}
         \includegraphics[scale = .3]{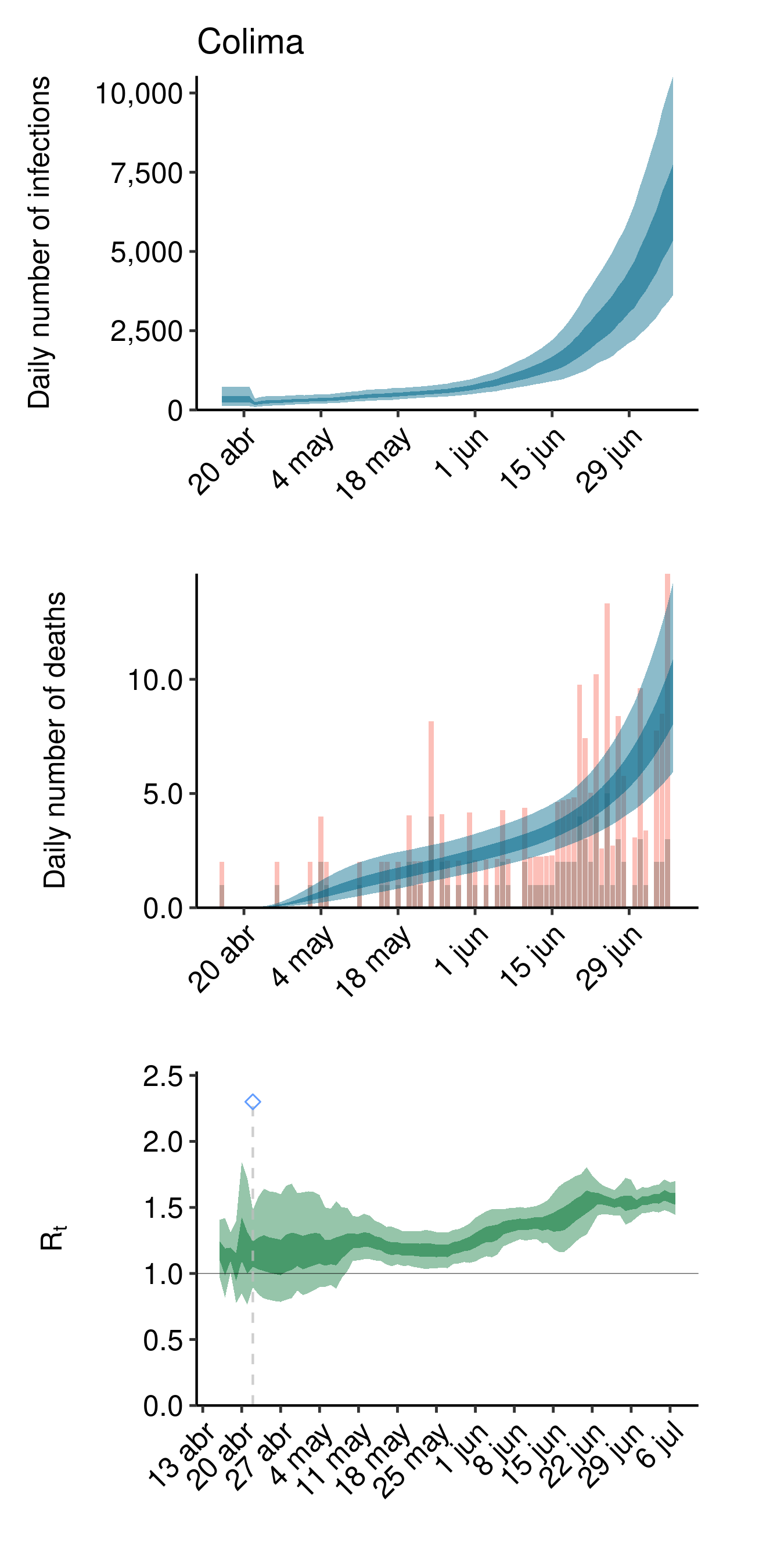}
         \includegraphics[scale = .3]{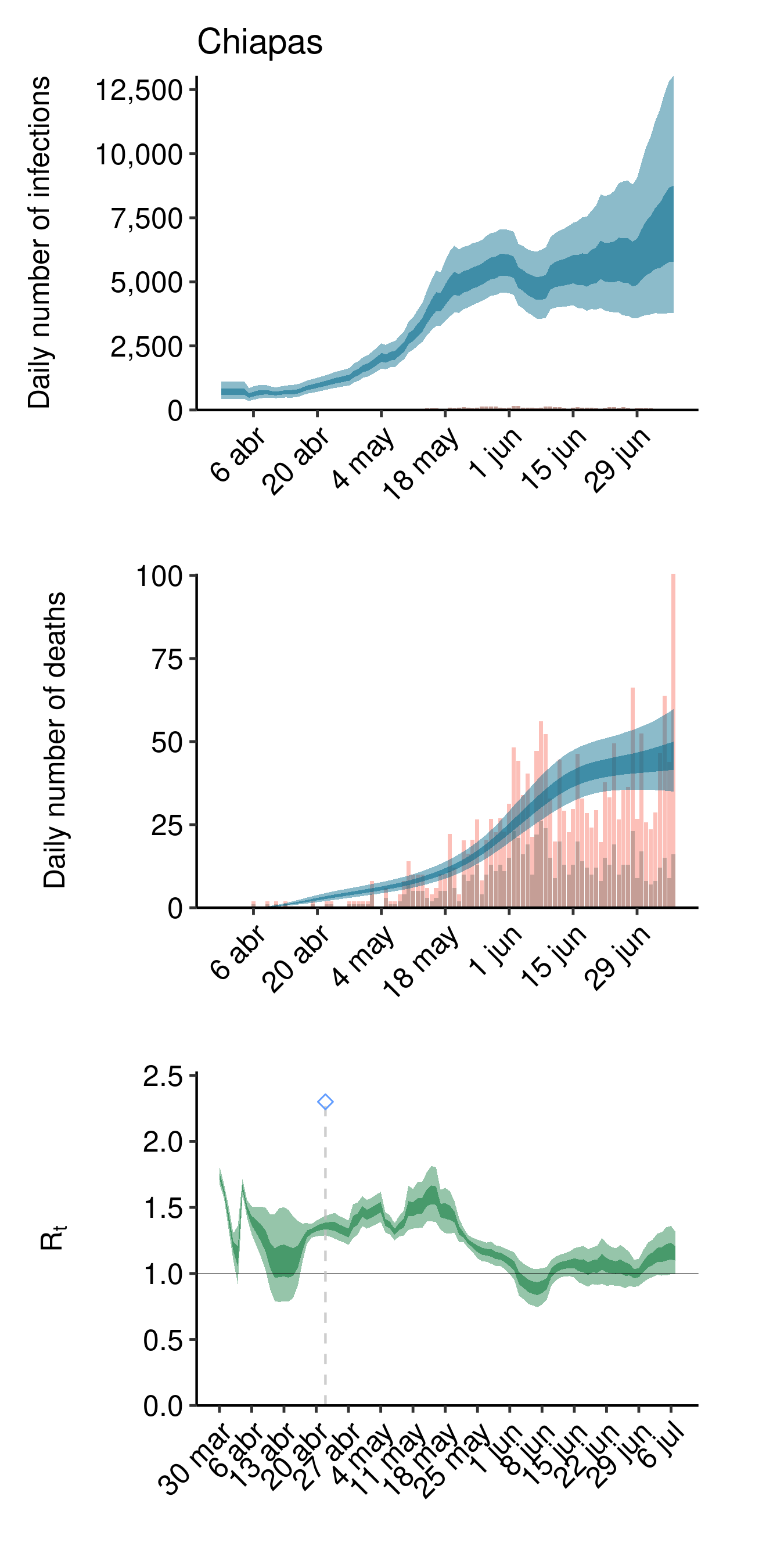}
        \includegraphics[scale = .3]{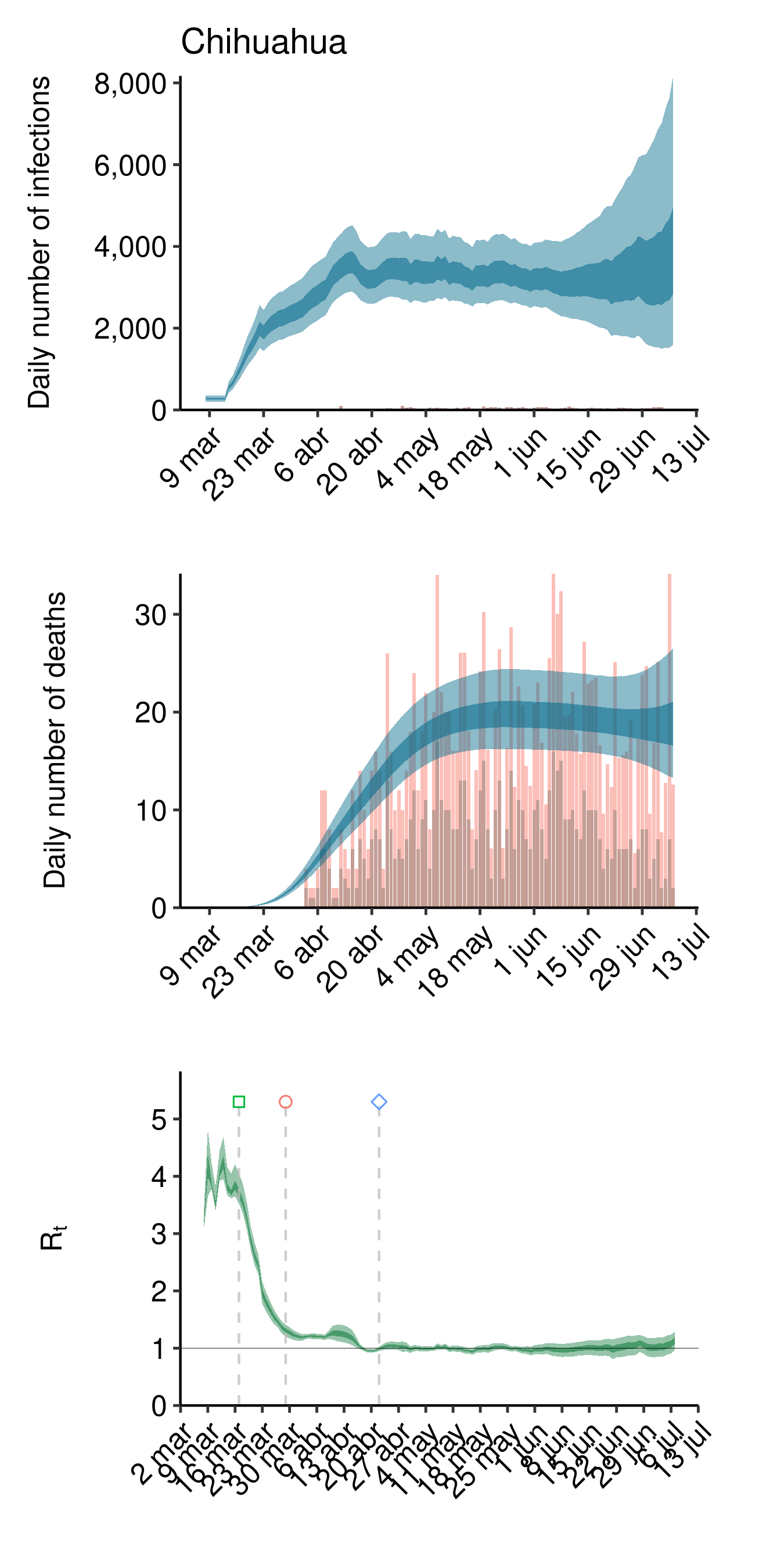}
    \end{subfigure}
    \medskip
    \begin{subfigure}{1.2\textwidth}
         \includegraphics[scale = .3]{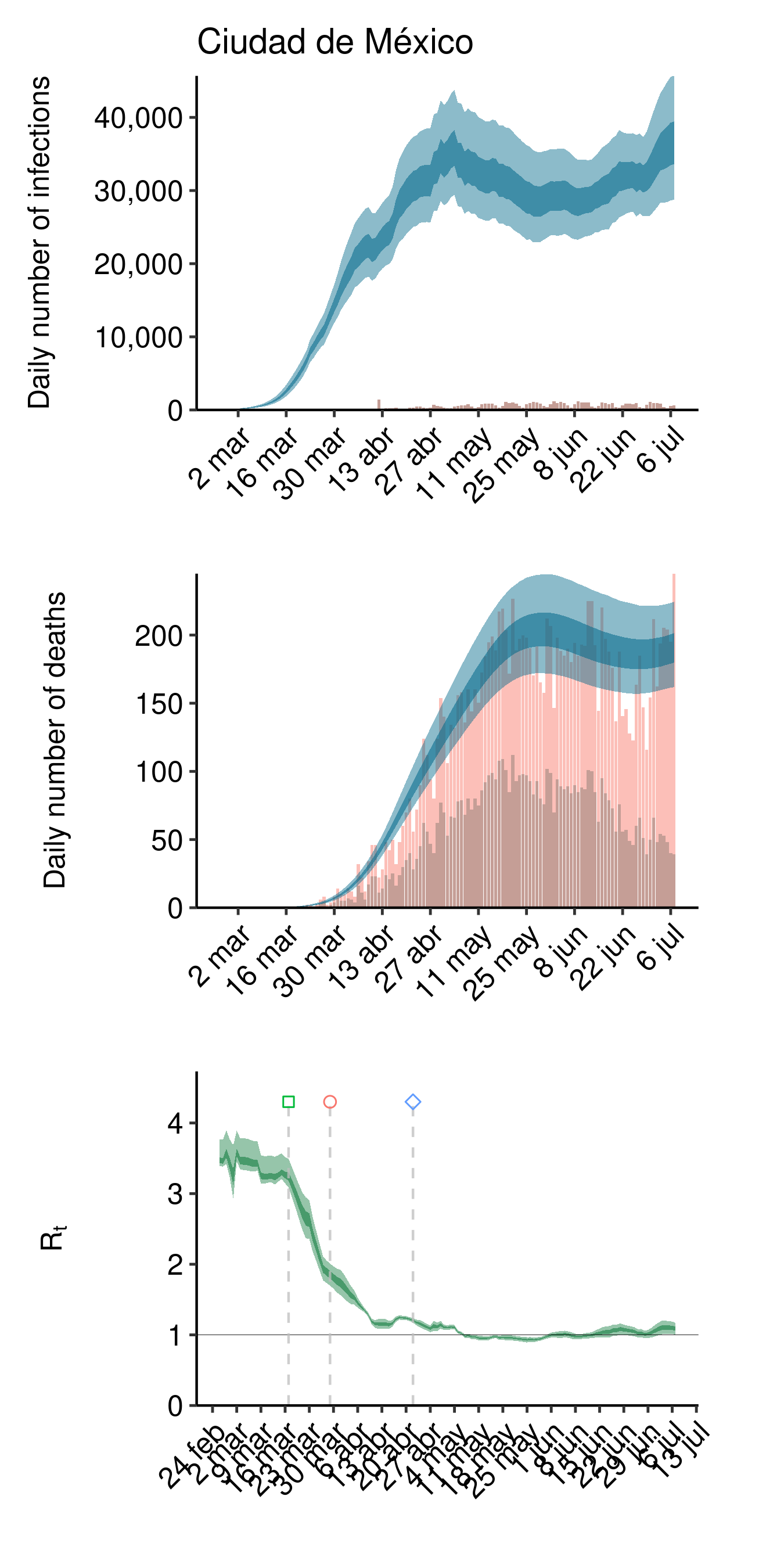}
        \includegraphics[scale = .3]{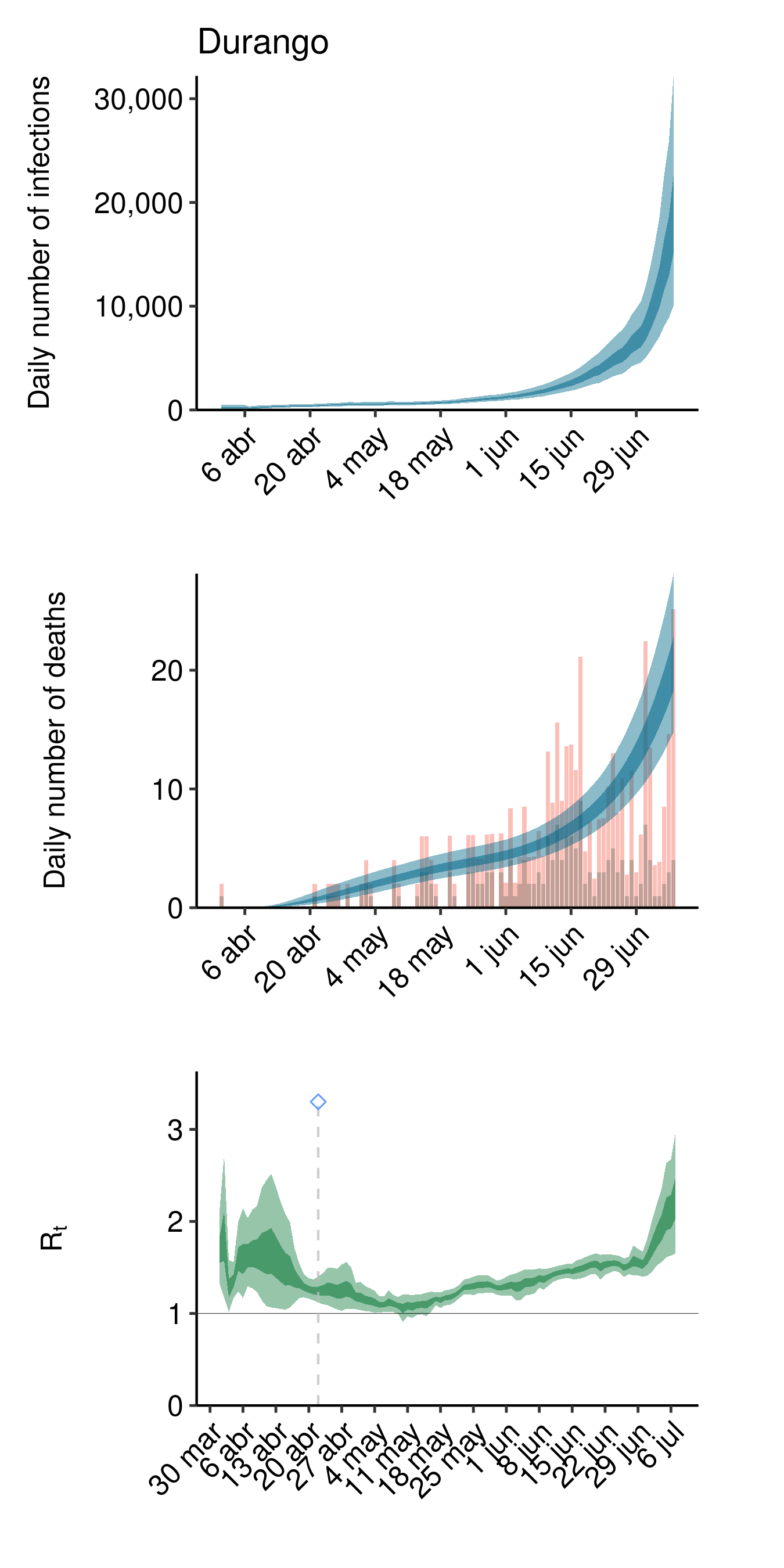}
        \includegraphics[scale = .3]{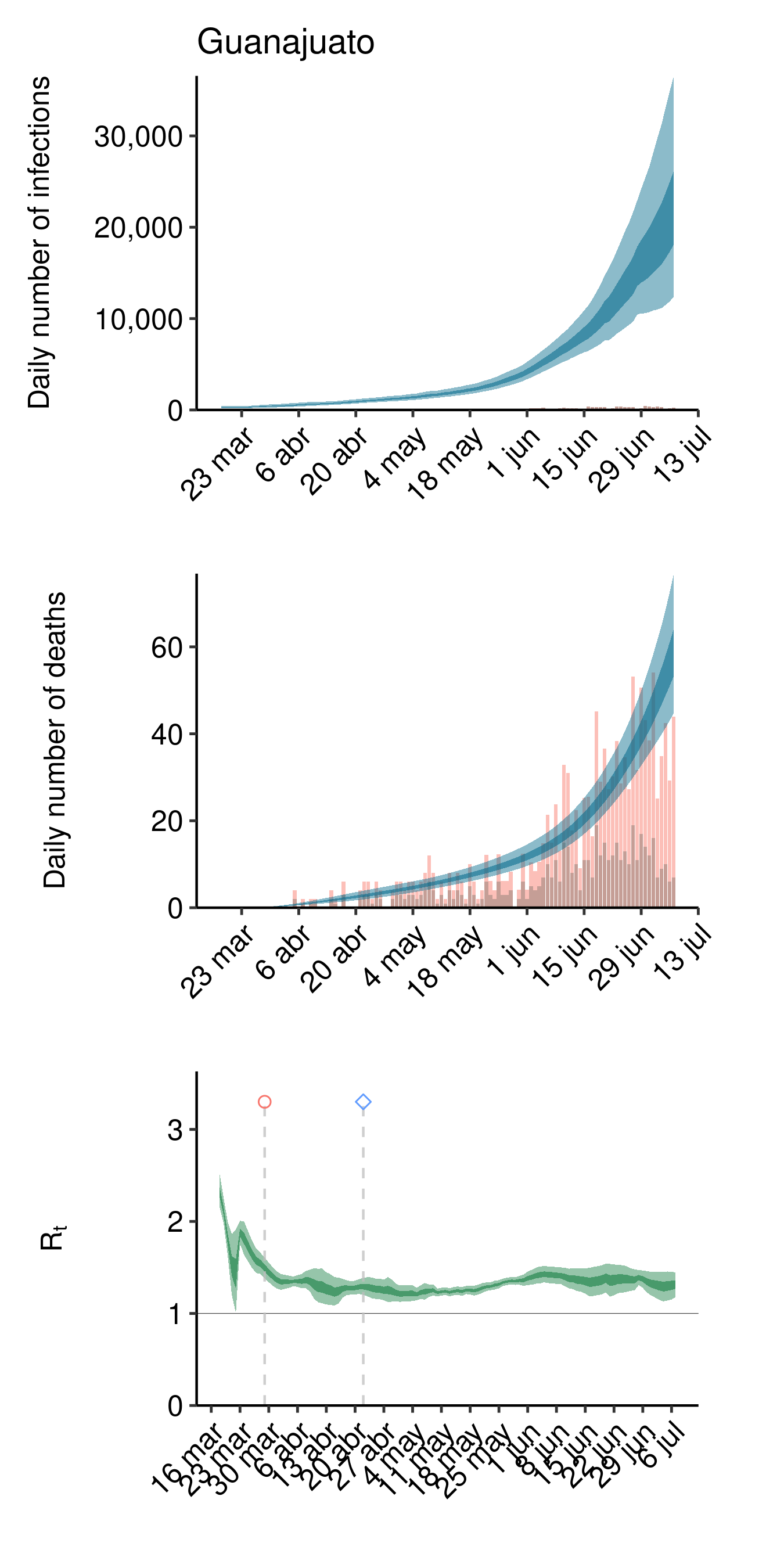}
         \includegraphics[scale = .3]{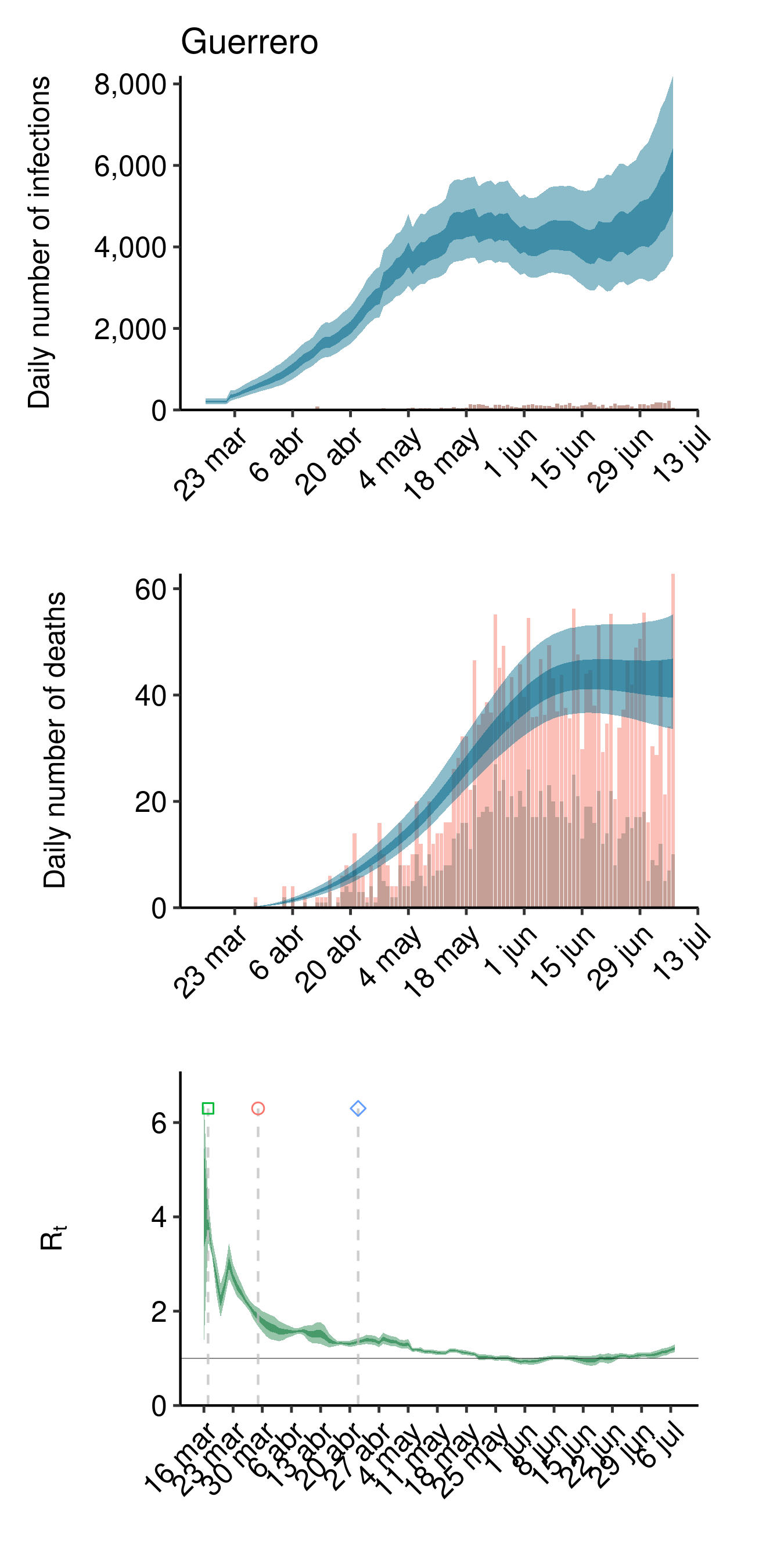}
    \end{subfigure}
\end{figure}%
\begin{figure}[ht]\ContinuedFloat
\centering
    \begin{subfigure}{1.2\textwidth}
         \includegraphics[scale = .3]{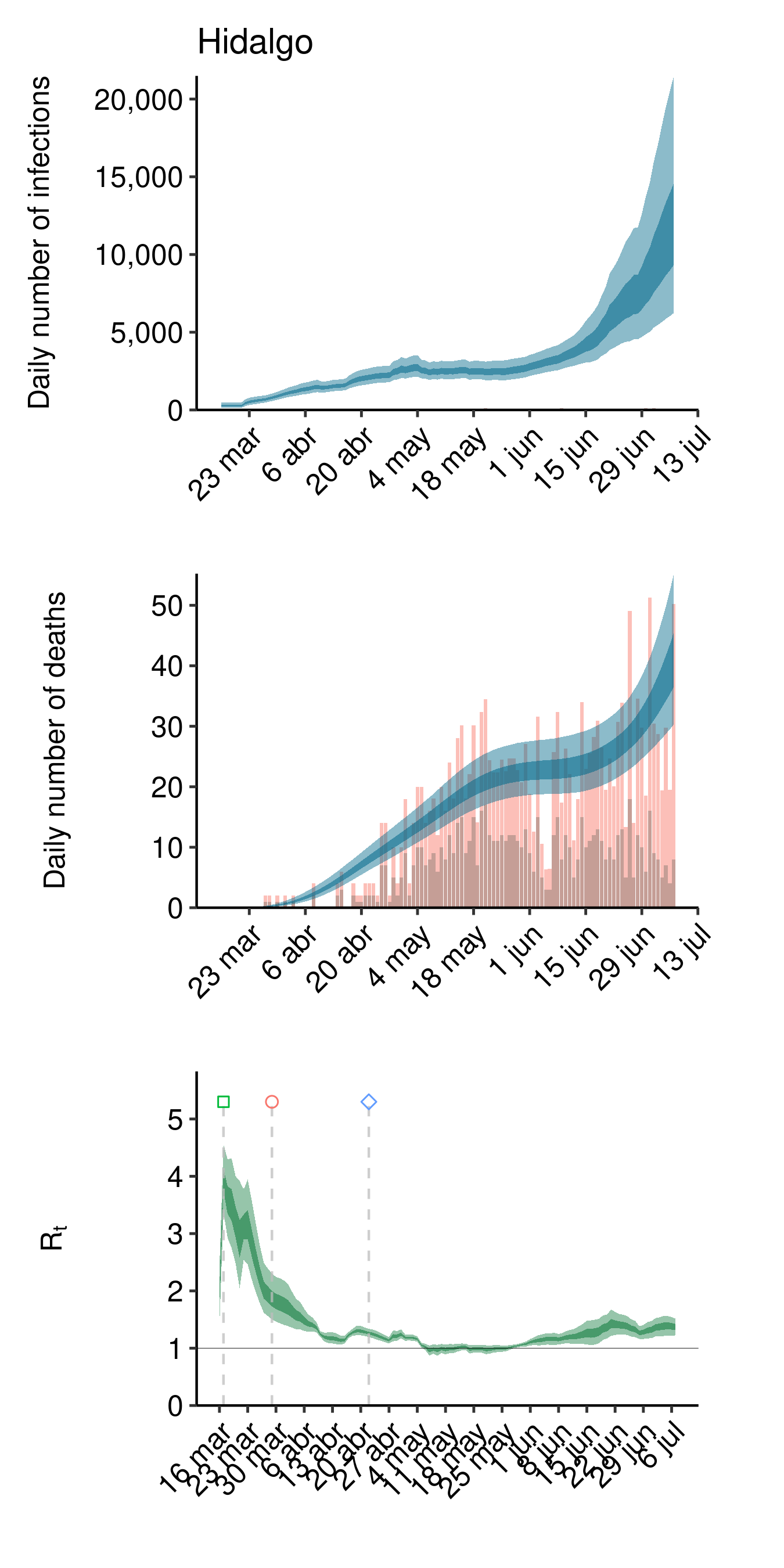}
        \includegraphics[scale = .3]{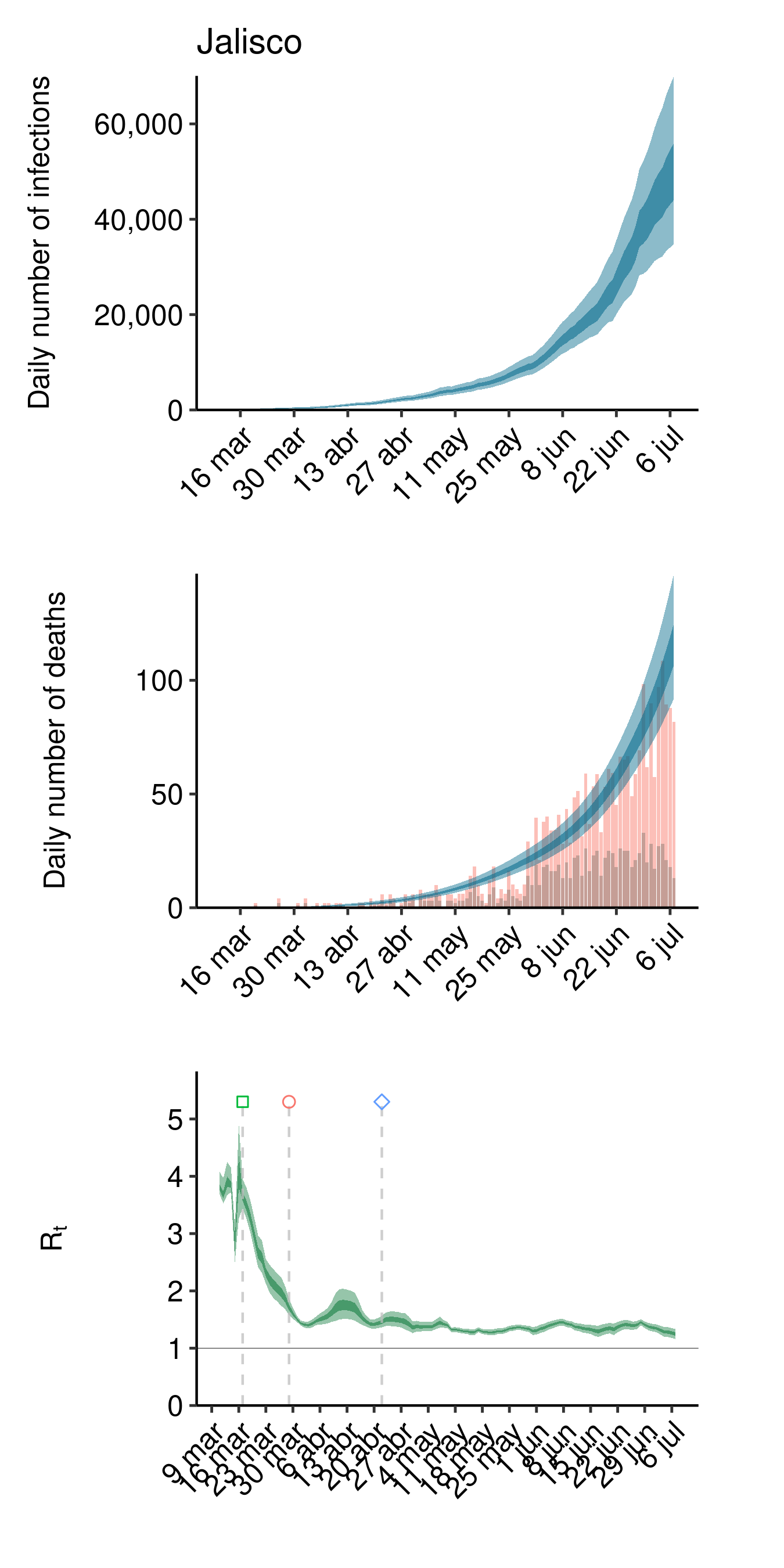}
         \includegraphics[scale = .3]{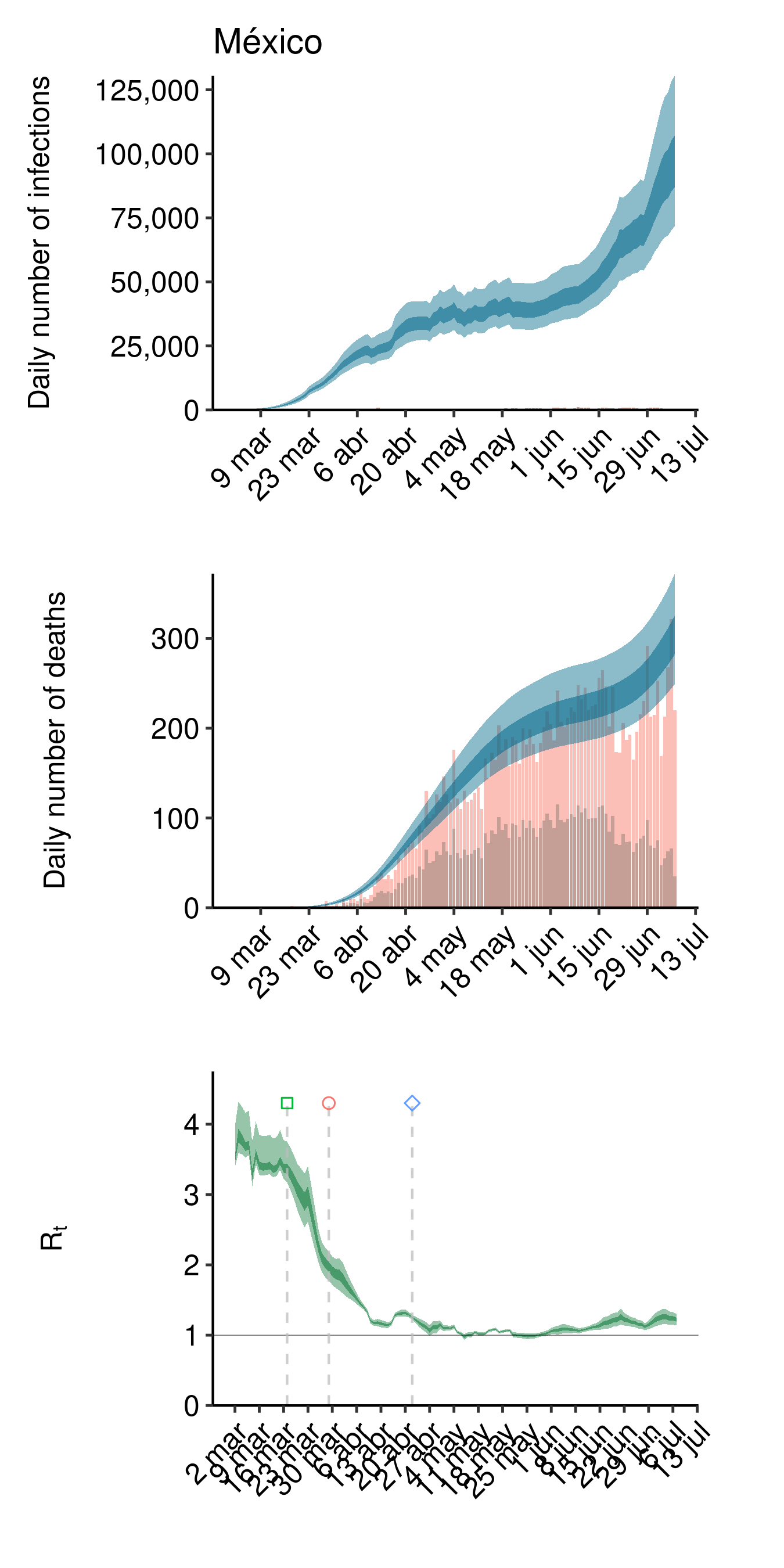}
		\includegraphics[scale = .3]{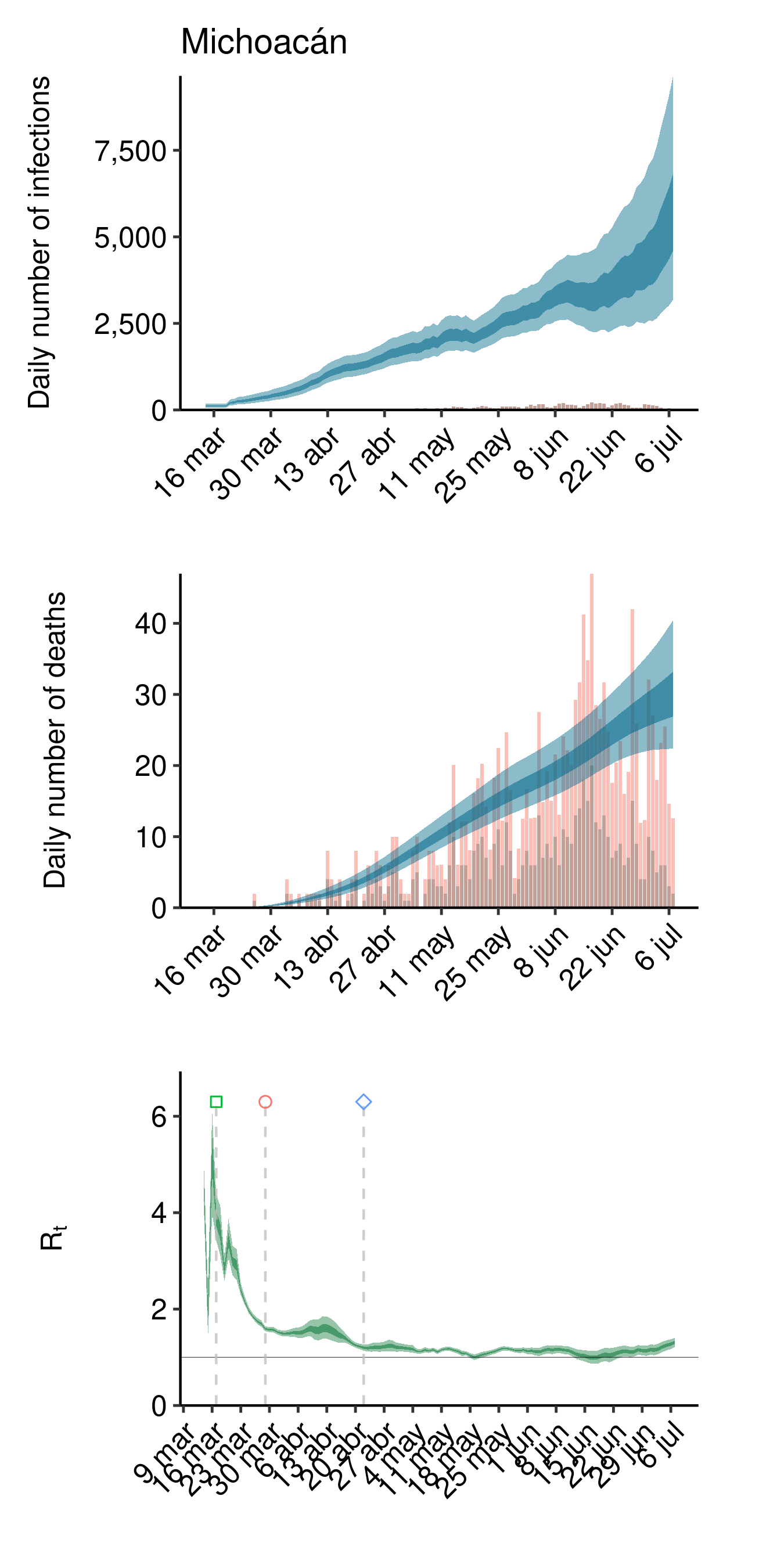}
       \end{subfigure}
    \medskip
    \begin{subfigure}{1.2\textwidth}
         \includegraphics[scale = .3]{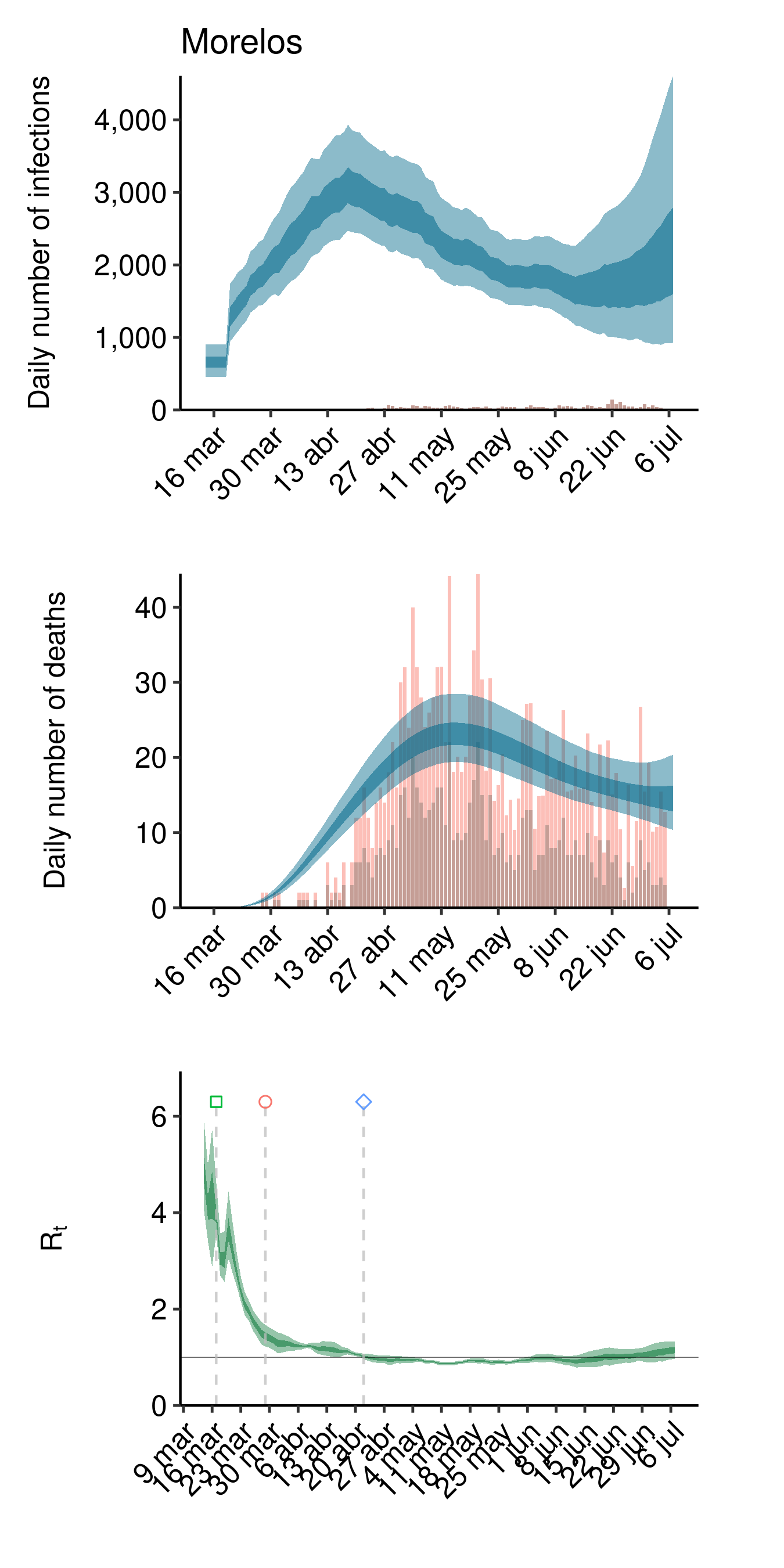}
         \includegraphics[scale = .3]{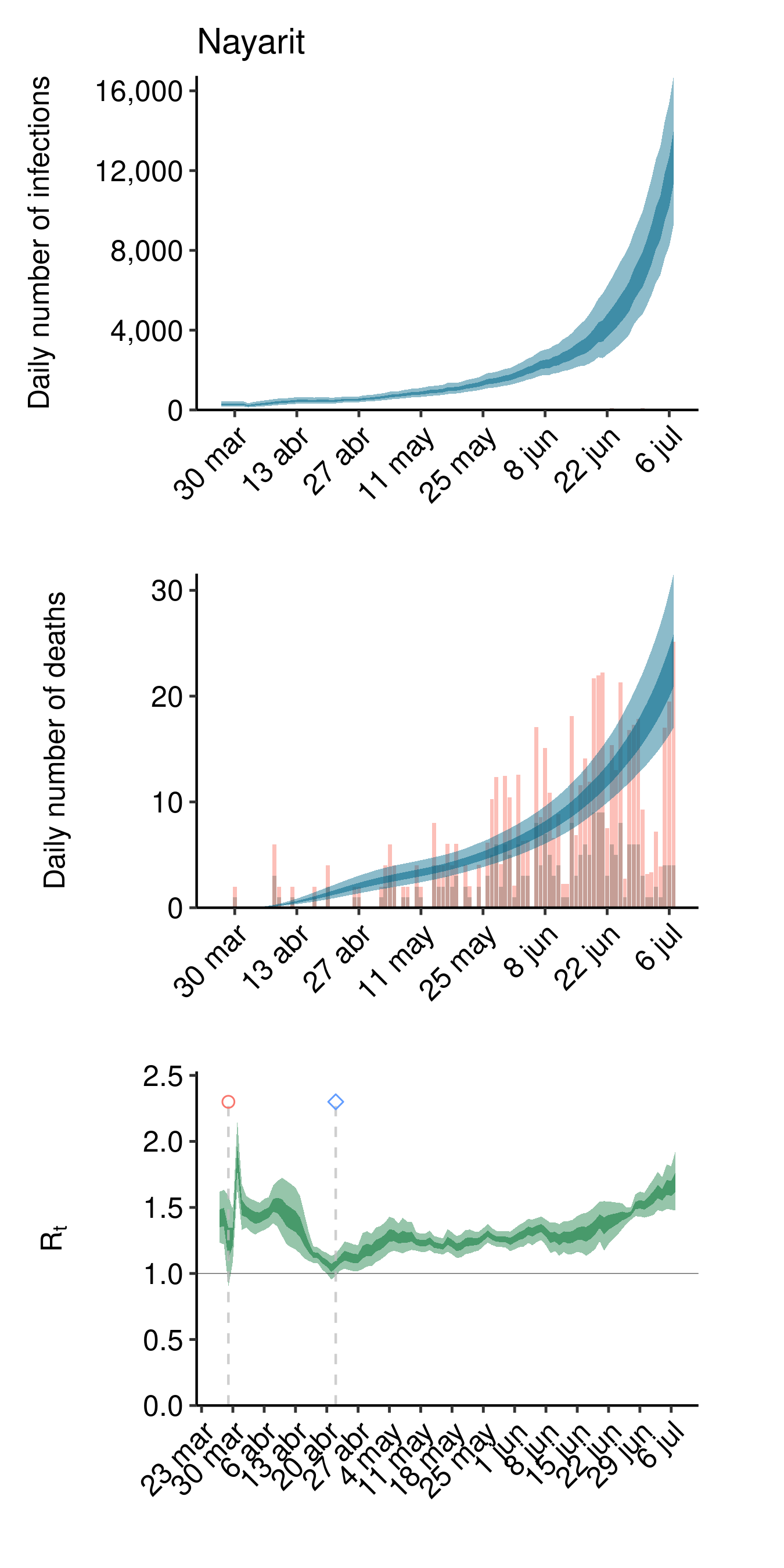}
          \includegraphics[scale = .3]{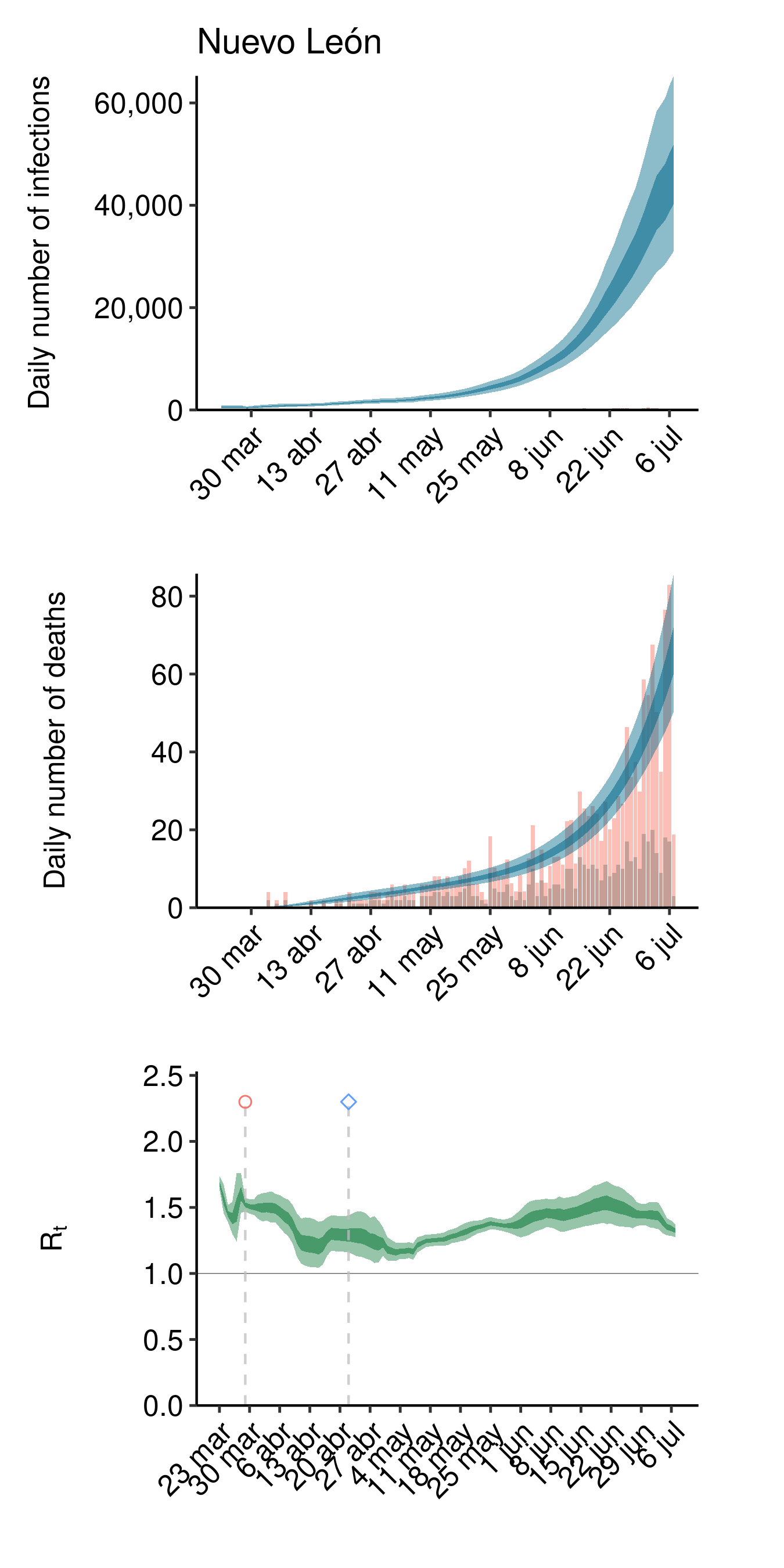}
        \includegraphics[scale = .3]{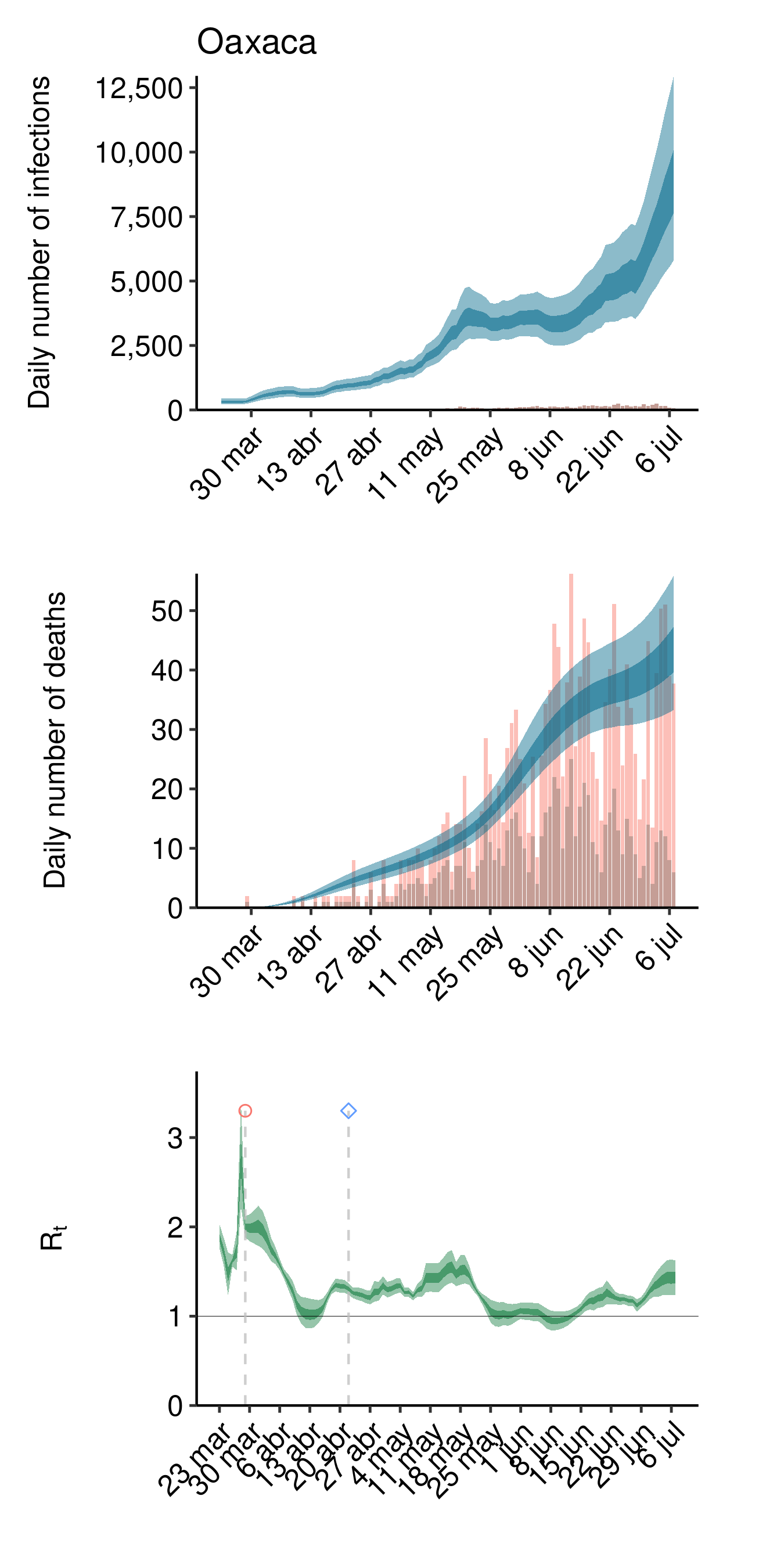}
    \end{subfigure}
	\medskip
    \begin{subfigure}{1.2\textwidth}
        \includegraphics[scale = .3]{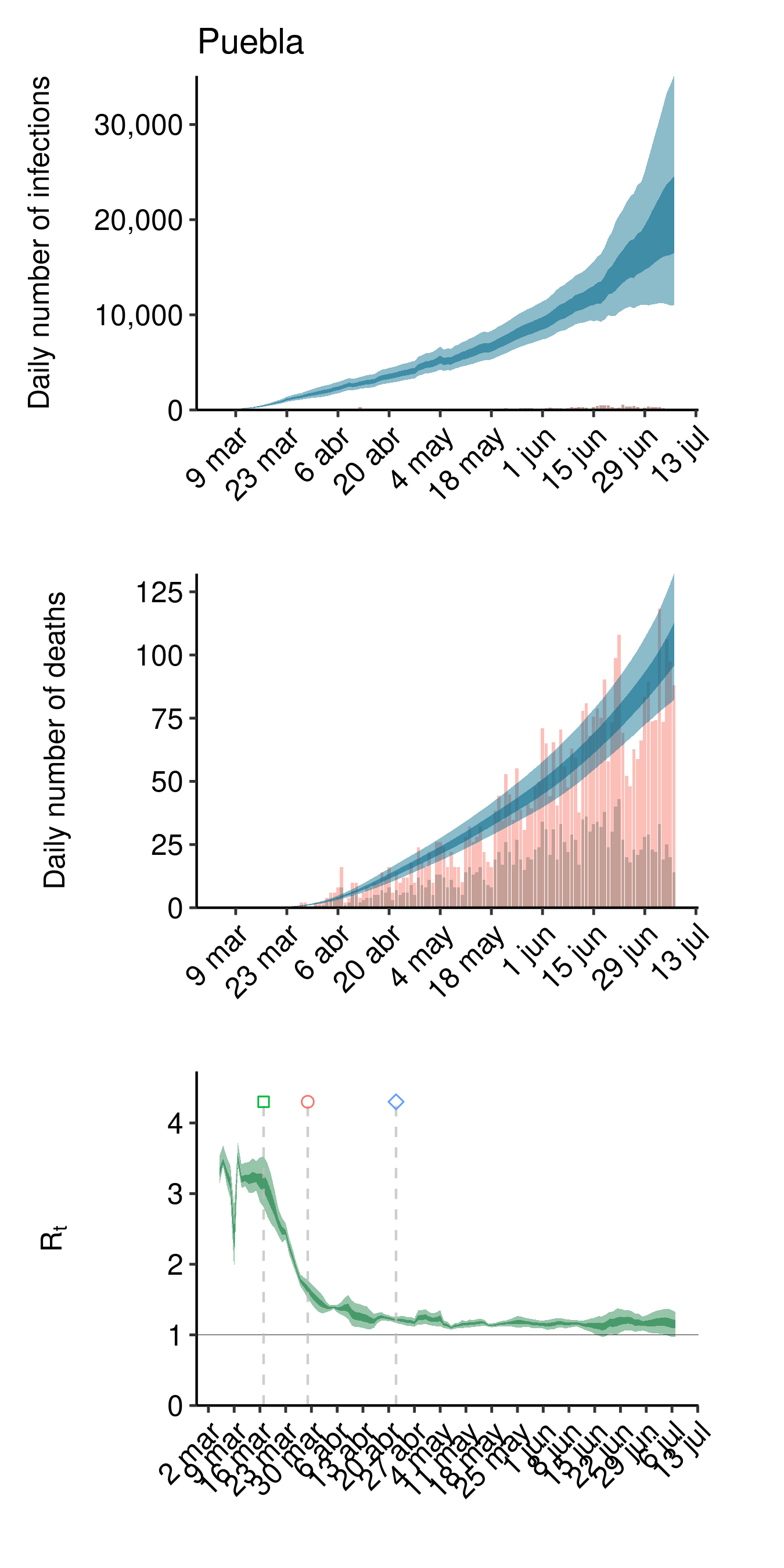}
		\includegraphics[scale = .3]{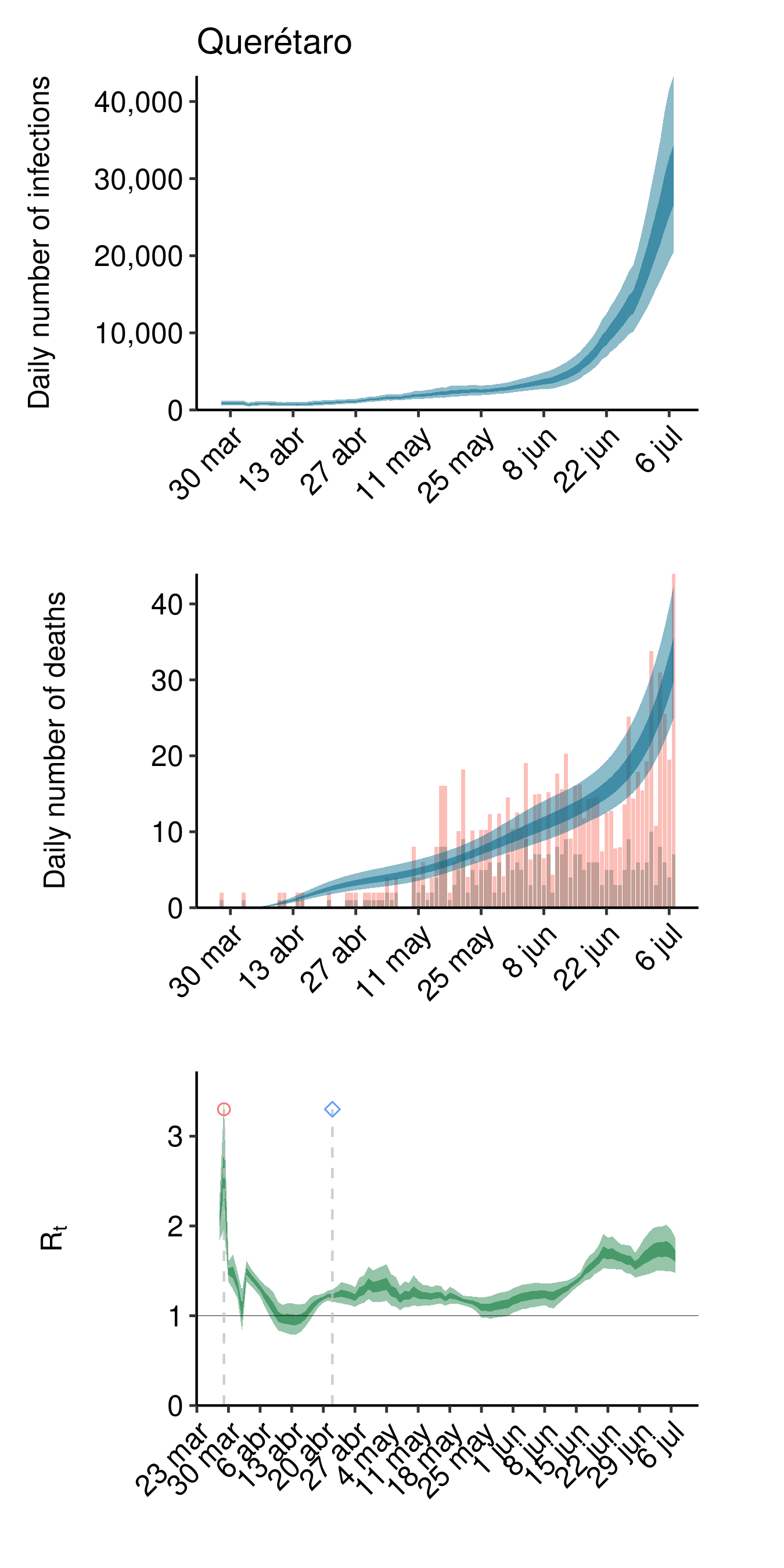}
        \includegraphics[scale = .3]{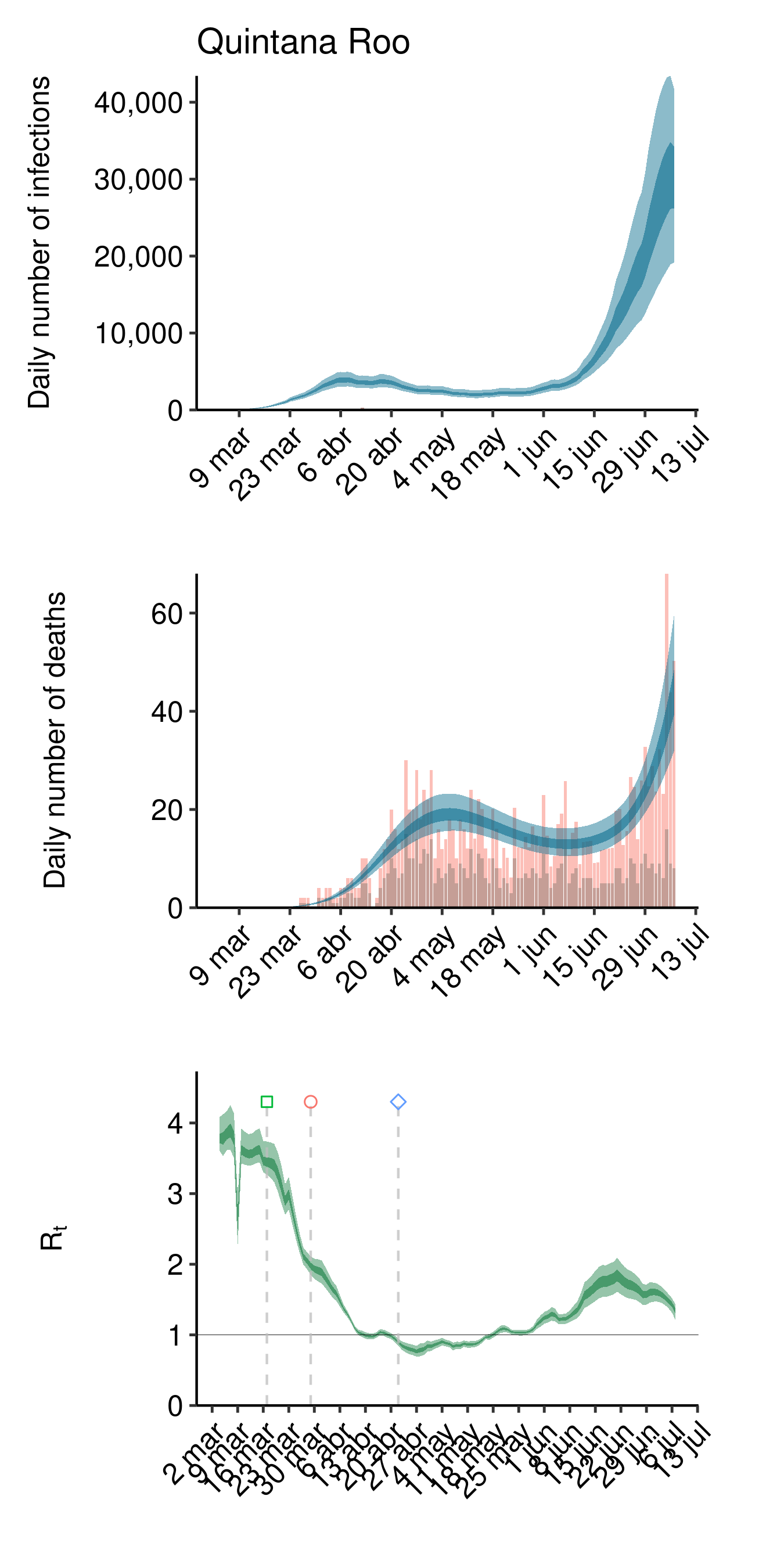}
         \includegraphics[scale = .3]{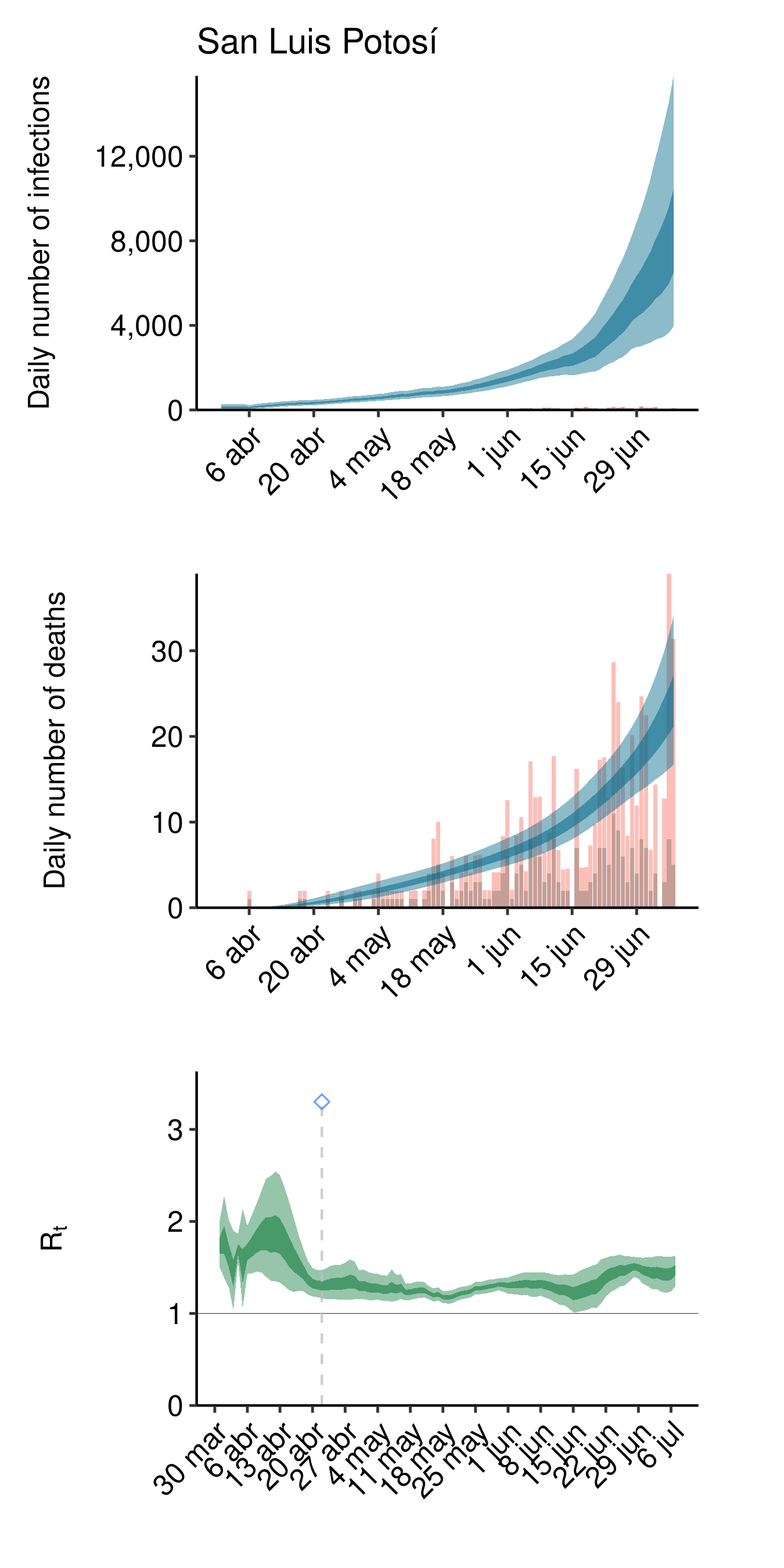}
    \end{subfigure}
\end{figure}%
\begin{figure}[ht]\ContinuedFloat
    \centering
    \begin{subfigure}{1.2\textwidth}
      	\includegraphics[scale = .3]{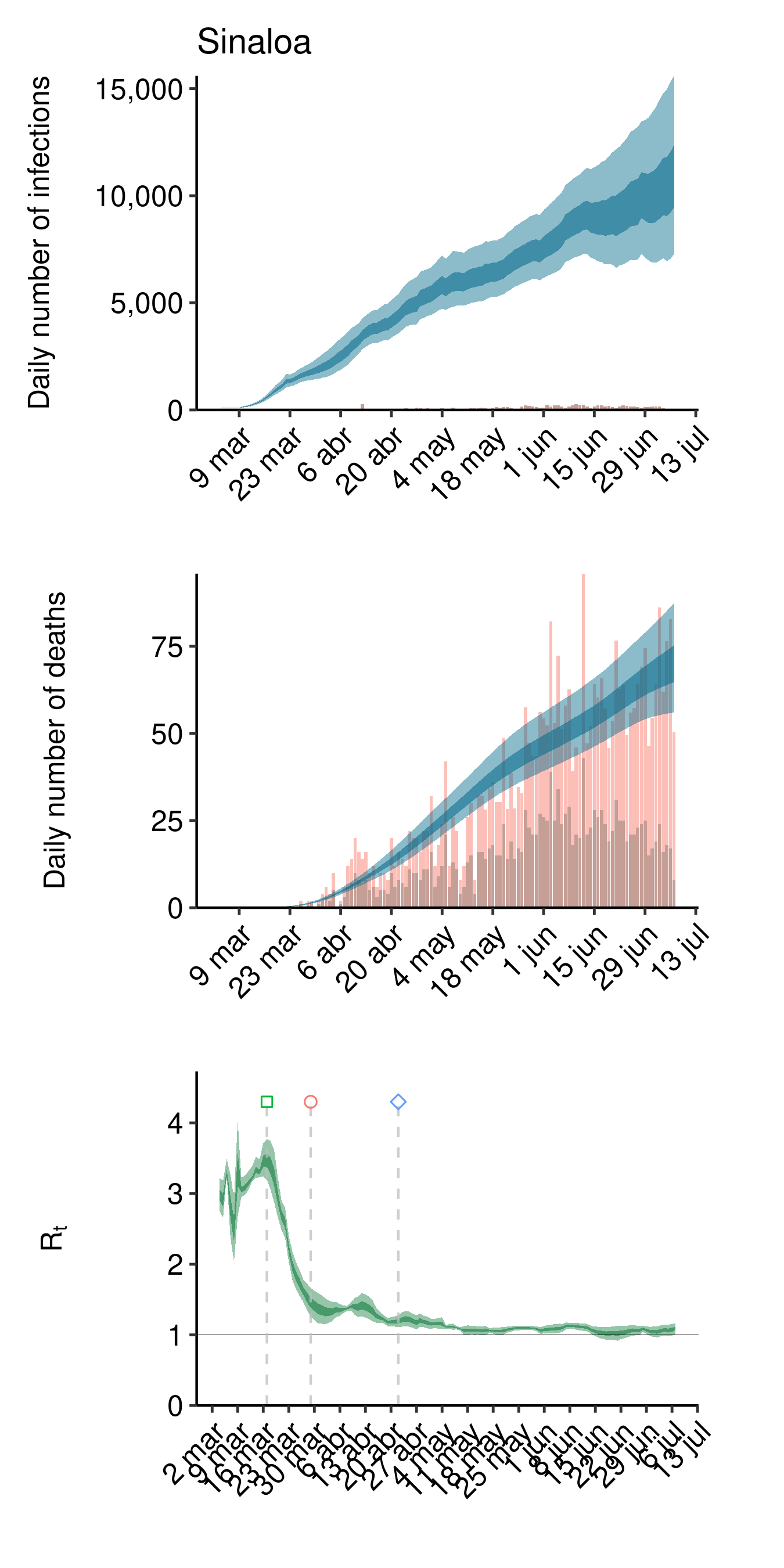}
      	\includegraphics[scale = .3]{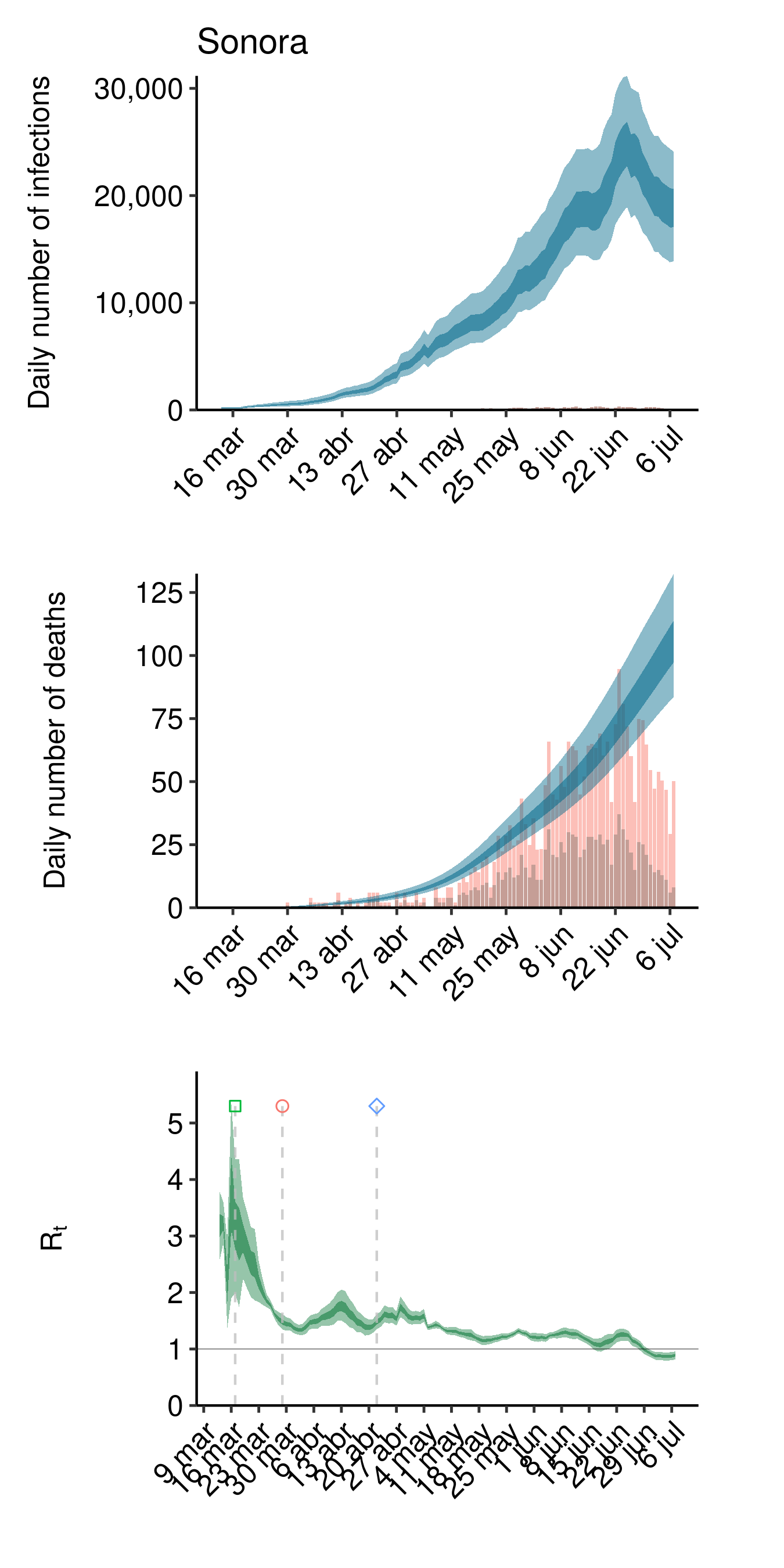}
        \includegraphics[scale = .3]{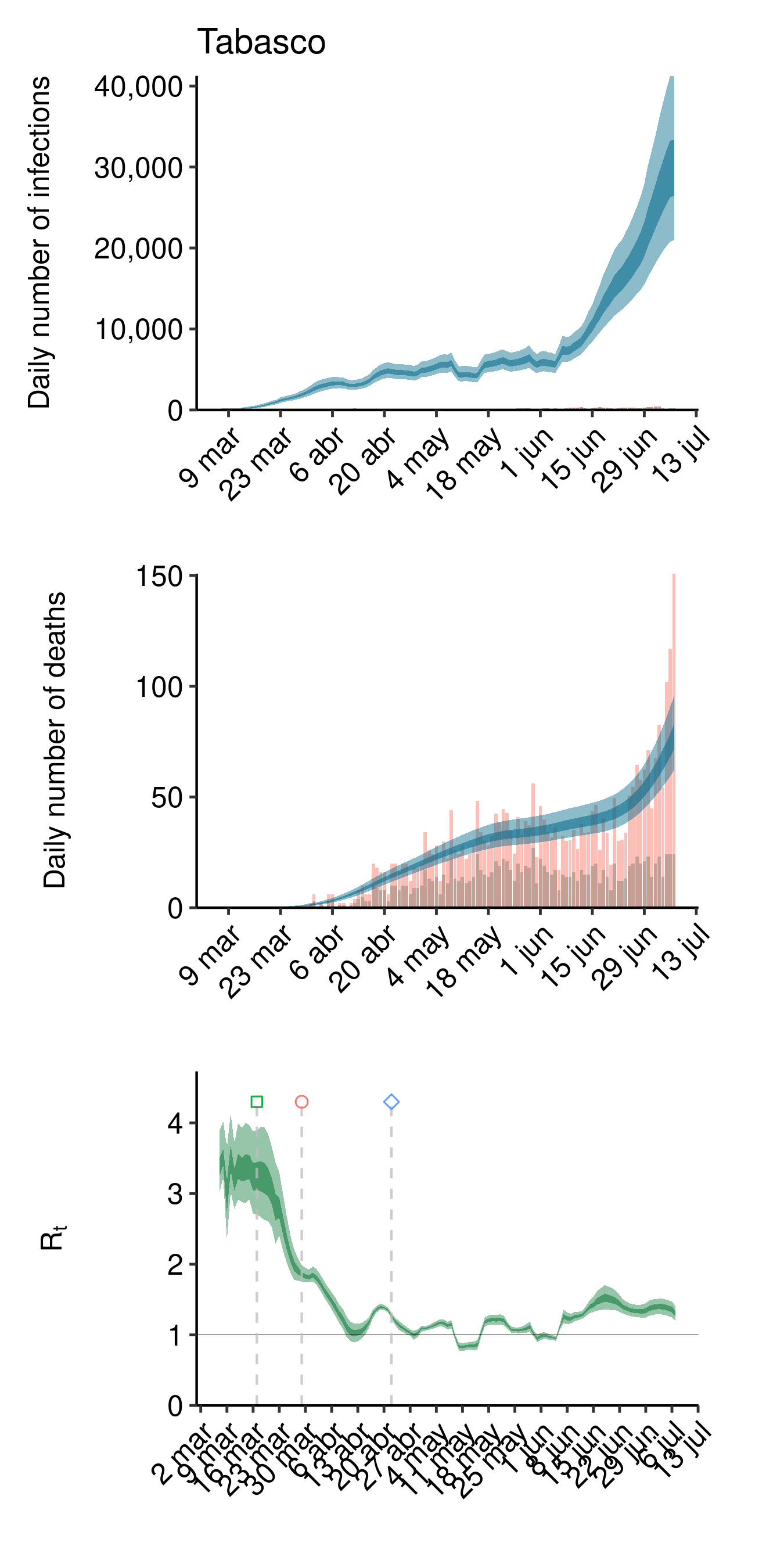}
        \includegraphics[scale = .3]{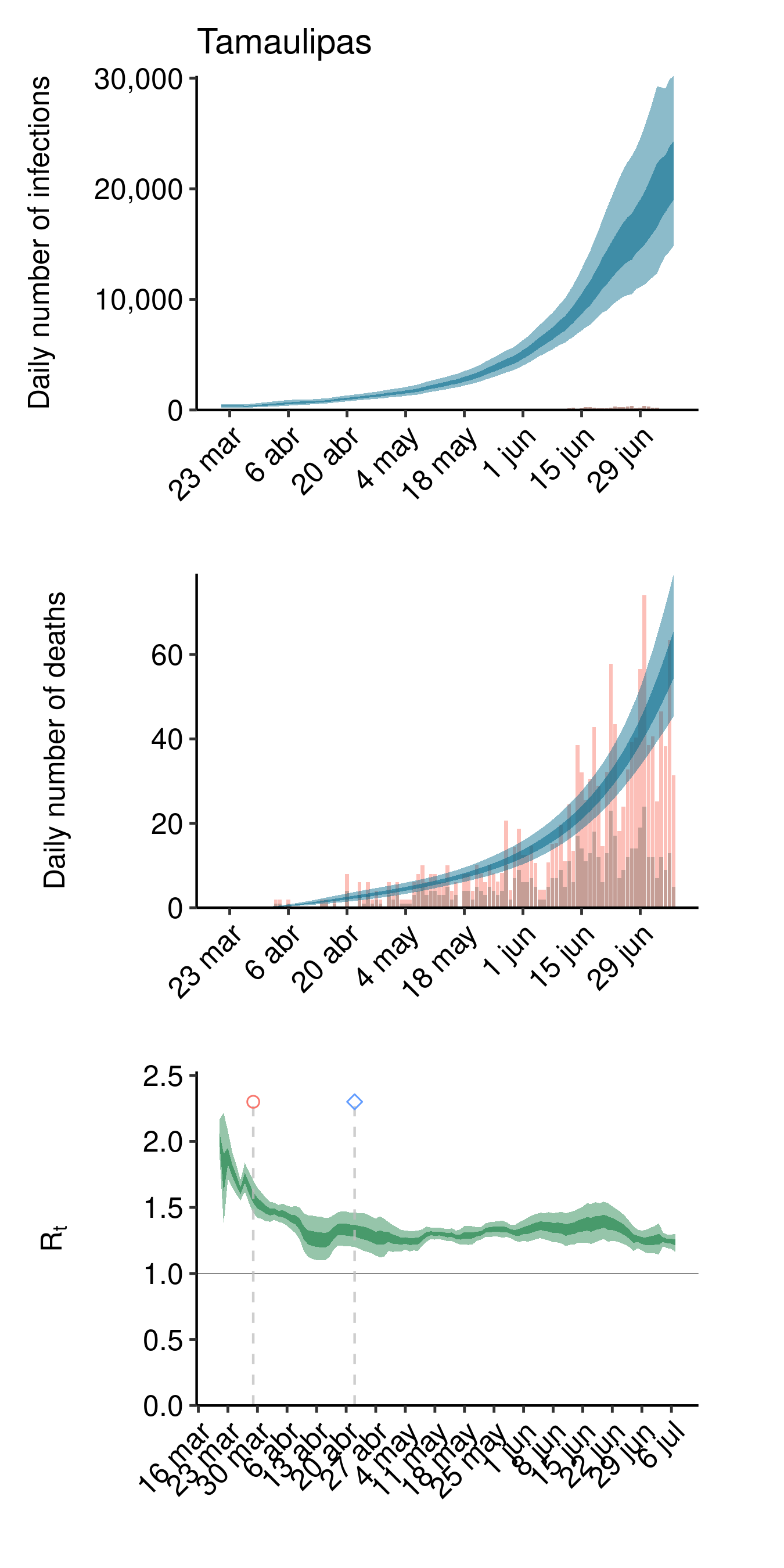}      
    \end{subfigure}
	\medskip
    \begin{subfigure}{1.2\textwidth}
         \includegraphics[scale = .3]{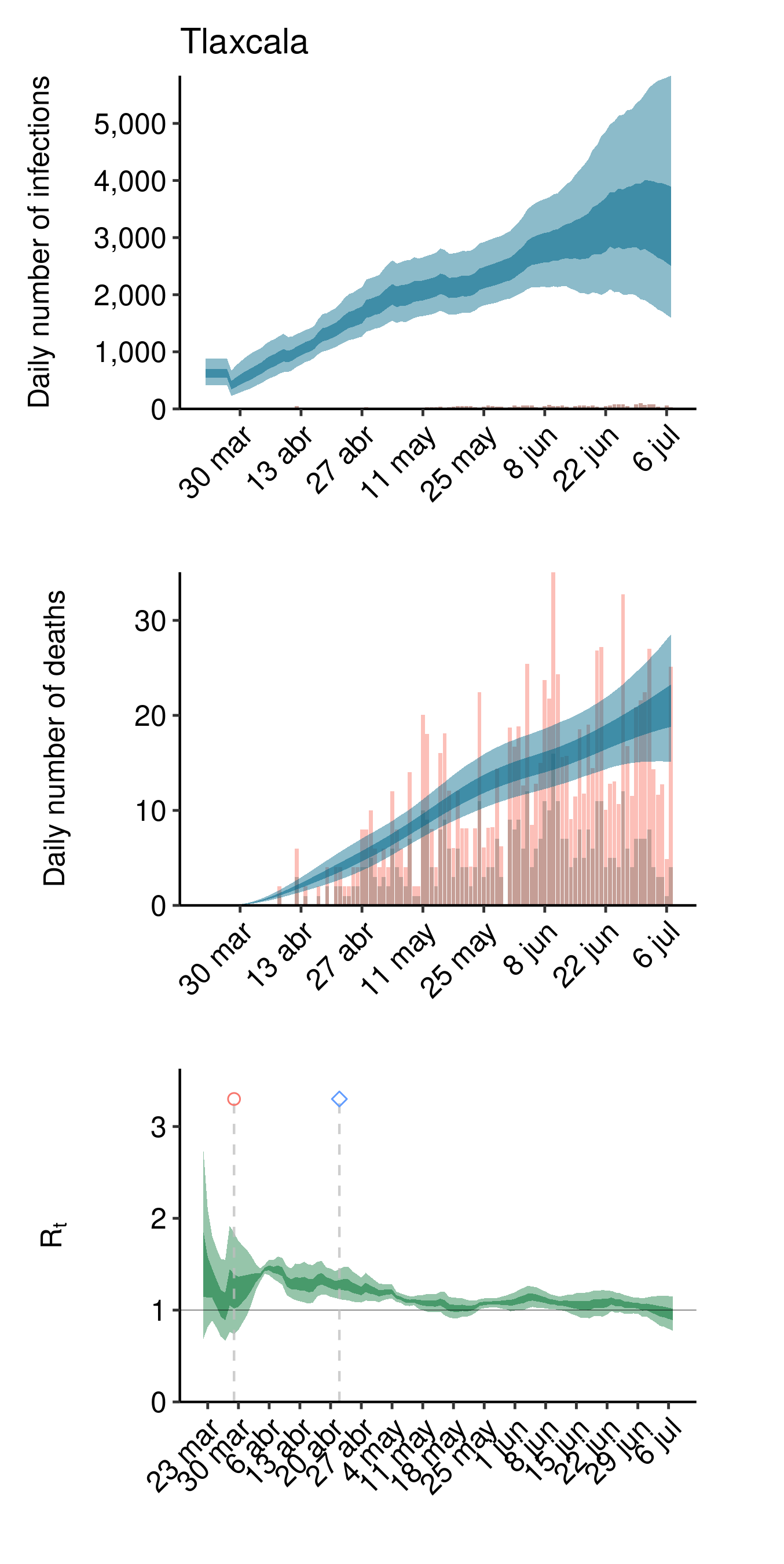}
         \includegraphics[scale = .3]{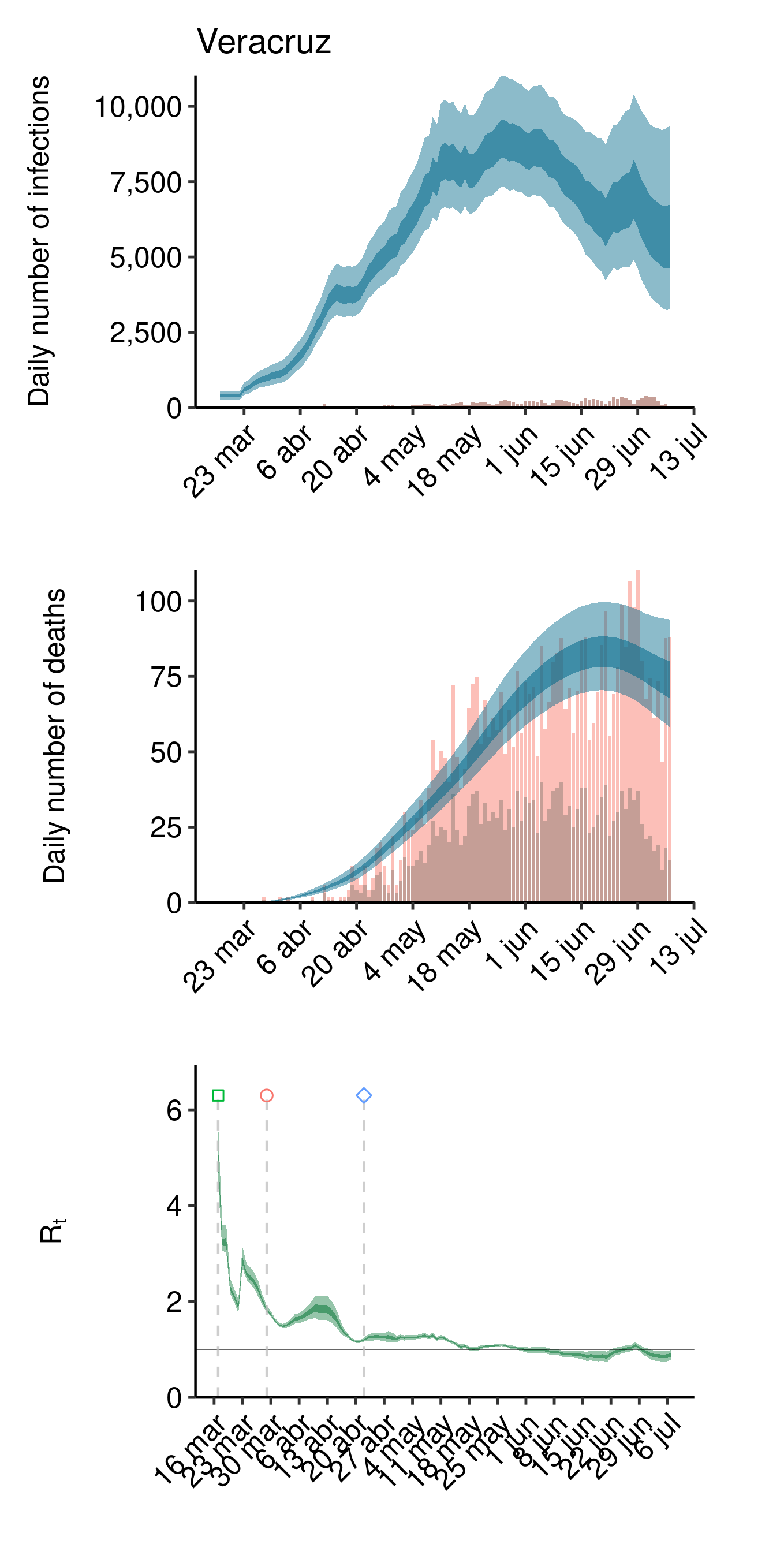}
         \includegraphics[scale = .3]{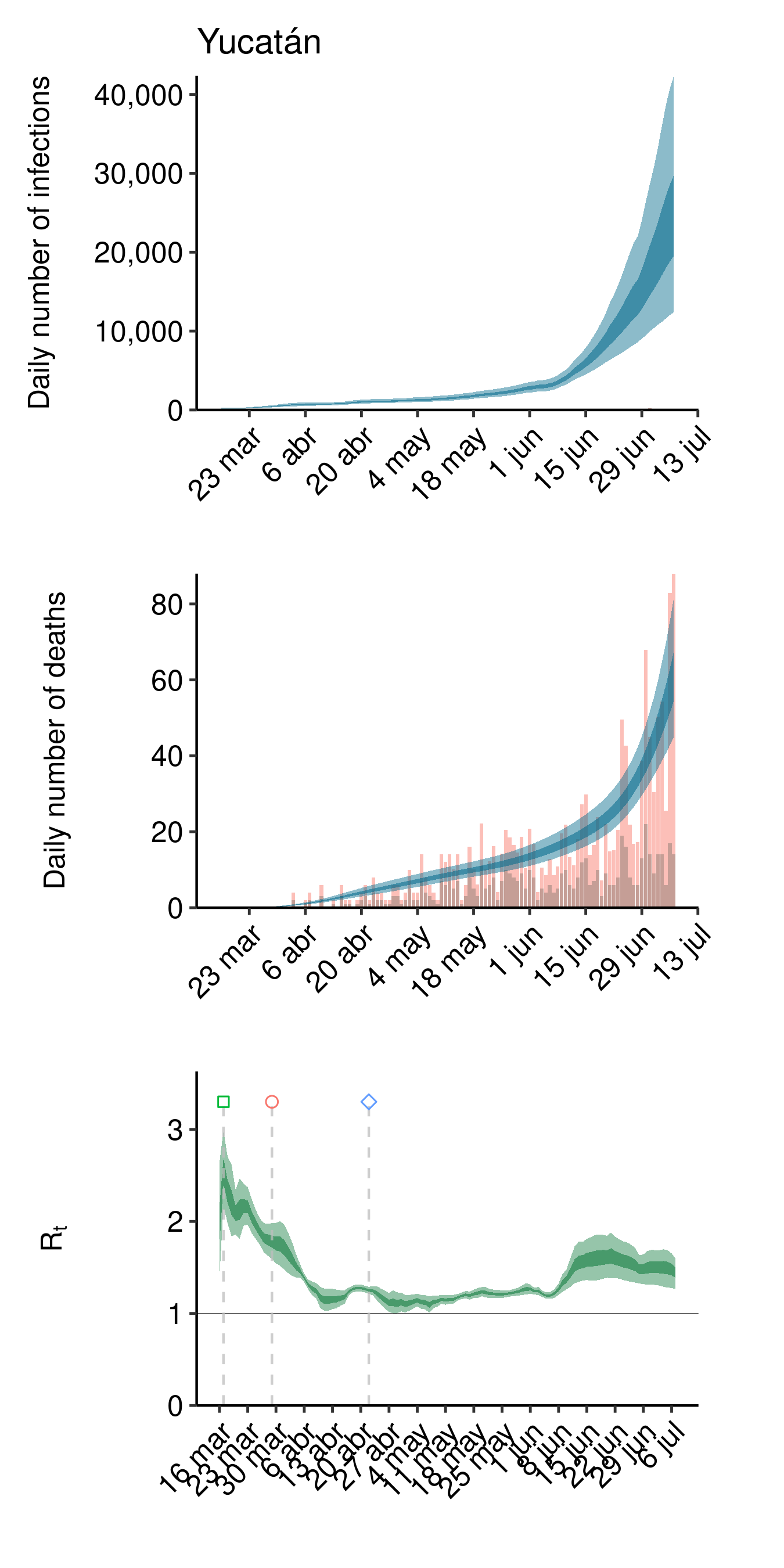}
        \includegraphics[scale = .3]{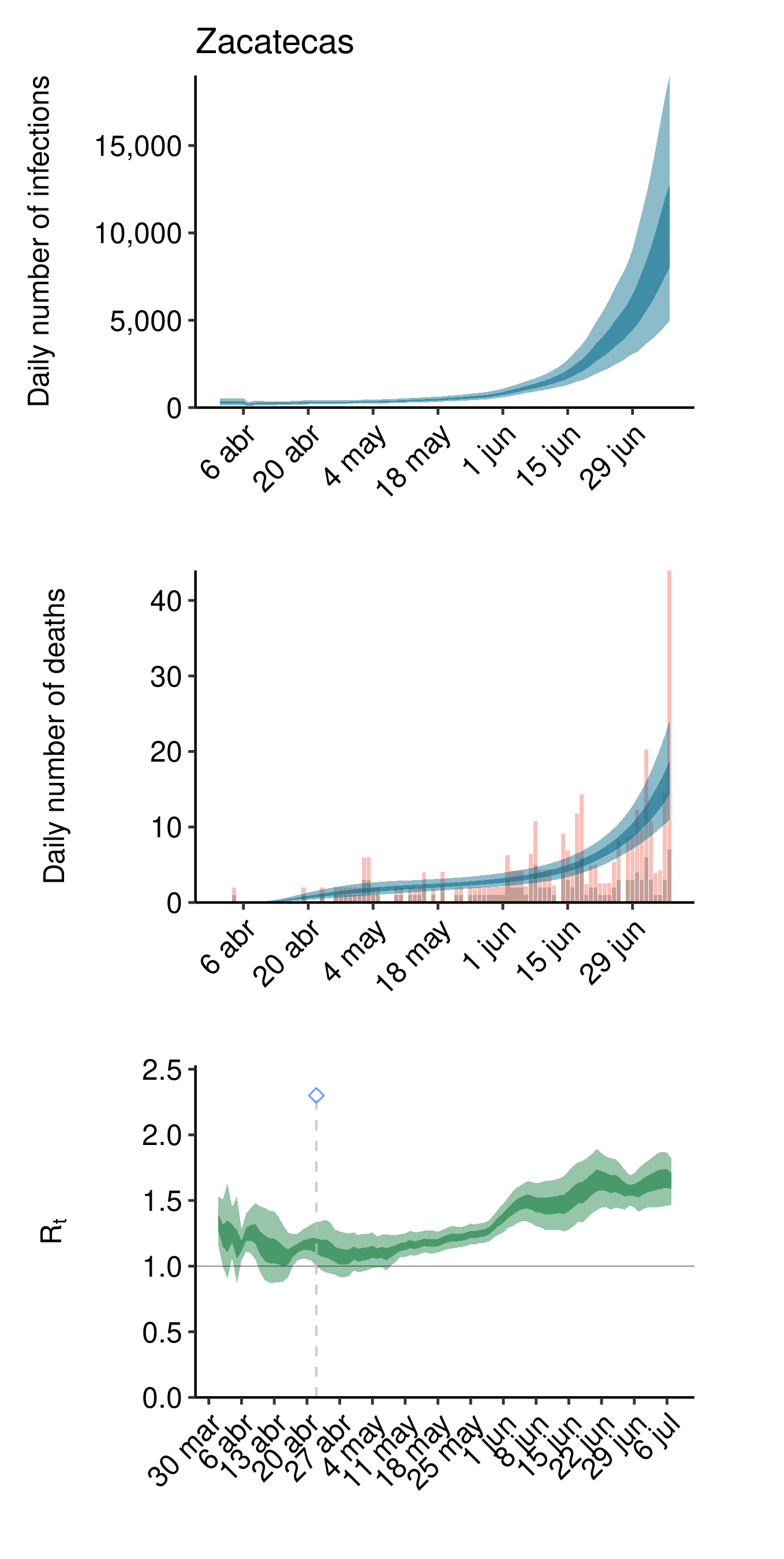}      
    \end{subfigure}
	
    \caption[]{For each state, the first plot has the daily number of infections, brown bars are reported cases, blue bands are predicted infections, dark blue 50\% CI, and light blue 95\% CI. The second plot shows daily number of deaths, brown bars are reported deaths, pink bars are estimated deaths adjusted for the reporting lag and the under-reporting, blue bands are predicted deaths, and CI as in the first plot. The third plot is the time-varying reproduction number, dark green 50\% CI, light green 95\% CI. The dotted line with a square is the date of school closures. The dotted line with the circle marks the start of sanitary emergency. Lastly, the line with the rhombus is the start of phase 3.}
    \label{fig:infections}
\end{figure}

\section{Discussion}\label{section:Discussion}

We applied the semi-mechanistic Bayesian hierarchical model found in \cite{flaxman2020report} as an alternative model to estimate the number of infections and the reproduction number in Mexico. In this context, we have incomplete and biased data. We expect the model to provide more accurate estimates, since the estimation is made with deaths observed over time, on the basis that the death data is more reliable than the data of infected people. Weekly updates can be checked at \url{https://covid19mex.netlify.app/}.

The results presented here suggest an ongoing epidemic in which substantial reductions in the average reproduction number have been achieved through non-pharmaceutical interventions. However, our results also show that so far the changes in mobility have not been enough to reduce the reproduction number below one. Therefore we predict continued growth of the epidemic across Mexico and increases in the associated number of cases and deaths unless further actions are taken.

Our results also reveal extensive heterogeneity in predicted attack rates between states, suggesting that the epidemic is at a far more advanced stage in some states compared to others. Despite this, in most states, a small proportion of individuals has been infected to date, indicating that herd immunity is not close yet. The exception is perhaps Baja California, where almost half of the people have already been infected.

It is now established that SARS-CoV-2 pre-existing immune reactivity exists to some degree in the general population \citep[see, e.g., ][]{sette2020pre}. Pre-existing cross-reactivity against COVID-19 in a fraction of the human population is relevant in our modelling, that considers at every time a population adjustment for susceptible individuals. This can be introduced naturally in the model and is worthy of further exploration when there is more information. Also, there is still a lot to be done regarding the adjustment in death-delayed reporting, we will be able to set a distribution to death-delay for each state when more data is available. 

To conclude, let us highlight that this model uses mobility to predict the rate of transmission, neglecting the potential effect of additional behavioural changes or interventions such as increased mask wearing, changes in age specific movement, testing and tracing.

\section*{Acknowledgements}

We thank Paulina Preciado from geek end and V\'ictor M. Garc\'ia for their support during the elaboration of this work.

\clearpage
\bibliographystyle{chicago}
\begin{footnotesize}
	\bibliography{Covid_mx}
\end{footnotesize}

\end{document}